\documentclass[a4paper,UKenglish,cleveref, autoref, thm-restate]{lipics-v2021}

\hideLIPIcs 

\title{Computing the LCP Array of a Labeled Graph} 

\author{Jarno N. {Alanko}}{University of Helsinki, Finland}{jarno.alanko@helsinki.fi}{https://orcid.org/0000-0002-8003-9225}{}
\author{Davide {Cenzato}}{Ca' Foscari University of Venice, Italy}{davide.cenzato@unive.it}{https://orcid.org/0000-0002-0098-3620}{}
\author{Nicola {Cotumaccio}}{University of Helsinki, Finland}{nicola.cotumaccio@helsinki.fi}{https://orcid.org/0000-0002-1402-5298}{}
\author{Sung-Hwan {Kim}}{Ca' Foscari University of Venice, Italy}{sunghwan.kim@unive.it}{https://orcid.org/0000-0002-1117-5020}{}
\author{Giovanni {Manzini}}{University of Pisa, Italy}{giovanni.manzini@unipi.it}{https://orcid.org/0000-0002-5047-0196}{}
\author{Nicola {Prezza}}{Ca' Foscari University of Venice, Italy}{nicola.prezza@unive.it}{https://orcid.org/0000-0003-3553-4953}{}

\funding{\textit{Jarno Alanko, Nicola Cotumaccio}: Funded by  the Helsinki Institute for Information Technology (HIIT). \textit{Davide Cenzato, Sung-Hwan Kim, Nicola Prezza}: Funded by the European Union (ERC, REGINDEX, 101039208). Views and opinions expressed are however those of the author(s) only and do not necessarily reflect those of the European Union or the European Research Council. Neither the European Union nor the granting authority can be held responsible for them. \textit{Giovanni Manzini}: Funded by the Italian Ministry of Health, POS 2014-2020, project ID T4-AN-07, CUP I53C22001300001, by INdAM-GNCS Project CUP E53C23001670001 and by PNRR ECS00000017 Tuscany Health Ecosystem, Spoke~6 CUP I53C22000780001.}

\authorrunning{J. Alanko, D. Cenzato, N. Cotumaccio, S.-H. Kim, G. Manzini, and N. Prezza} 

\Copyright{Jarno Alanko, Davide Cenzato, Nicola Cotumaccio, Sung-Hwan Kim, Giovani Manzini, Nicola Prezza} 



\category{} 

\relatedversion{} 

\nolinenumbers 

\usepackage{soul}
\usepackage[utf8]{inputenc}
\usepackage[table,dvipsnames]{xcolor}
 \usepackage{cancel} 
\usepackage{float}
\usepackage{graphicx}
\usepackage{xcolor}
\usepackage{subcaption}
\usepackage[sort,nocompress]{cite}
\usepackage{todonotes}
\usepackage{verbatim}
\usepackage{subcaption}
\usepackage[normalem]{ulem}
\usepackage{array}
\usepackage{mathrsfs}
 \usepackage{multirow}
 \usepackage{blindtext}
\usepackage{hyperref}
\usepackage{graphicx}
\usepackage{amsmath}
\usepackage{mathtools}
\usepackage{mathrsfs}
 \usepackage{multirow}
 \usepackage{blindtext}
\usepackage{hyperref}
\usepackage{graphicx}
\usepackage{algorithm}
\usepackage{algpseudocode}

\usepackage{pgf}
\usepackage{tikz}
\usetikzlibrary{calc}
\usepackage{tikz-cd}
\tikzcdset{scale cd/.style={every label/.append style={scale=#1},
    cells={nodes={scale=#1}}}}

\usetikzlibrary{arrows,automata,positioning,matrix}
\newcommand{\LCP}{\mathsf{LCP}}
\newcommand{\LCPx}{\mathsf{LCP}^*}
\newcommand{\lcp}{\mathop{\mathsf{lcp}}}

\newcommand{\Ginfsup}{G_{is}}

\newcommand{\IN}[1]{\mathrm{in}_{#1}}
\newcommand{\OUT}[1]{\mathrm{out}_{#1}}

\newcommand{\notaNC}[1]{\textcolor{blue}{#1 NC}}
\newcommand{\notaNP}[1]{\textcolor{olive}{#1 NP}}

\newcommand{\notaSK}[1]{\textcolor{brown}{#1 SK}}
\newcommand{\notaJA}[1]{\textcolor{orange}{#1 JA}}

\EventEditors{John Q. Open and Joan R. Access}
\EventNoEds{2}
\EventLongTitle{42nd Conference on Very Important Topics (CVIT 2016)}
\EventShortTitle{CVIT 2016}
\EventAcronym{CVIT}
\EventYear{2016}
\EventDate{December 24--27, 2016}
\EventLocation{Little Whinging, United Kingdom}
\EventLogo{}
\SeriesVolume{42}
\ArticleNo{23}

\begin{document}

\maketitle

\begin{abstract}
The LCP array is an important tool in stringology, allowing to speed up pattern matching algorithms and enabling compact representations of the suffix tree. Recently, Conte \textit{et al.} [DCC 2023] and Cotumaccio \textit{et al.} [SPIRE 2023] extended the definition of this array to Wheeler DFAs and, ultimately, to arbitrary labeled graphs, proving that it can be used to efficiently solve matching statistics queries on the graph's paths.
In this paper, we provide the first efficient algorithm building the LCP array of a directed labeled graph with $n$ nodes and $m$ edges labeled over an alphabet of size $\sigma$.
The first step is to transform the input graph $G$ into a deterministic Wheeler pseudoforest $\Ginfsup$ with $O(n)$ edges encoding the lexicographically- smallest and largest strings entering in each node of the original graph. 
Using state-of-the-art algorithms, this step runs in $O(\min\{m\log n, m+n^2\})$ time on arbitrary labeled graphs, and in $O(m)$ time on Wheeler DFAs.
The LCP array of $G$ stores the longest common prefixes between those strings, i.e. it can easily be derived from the LCP array of $\Ginfsup$.
After arguing that the natural generalization of a compact-space LCP-construction algorithm by Beller \textit{et al.} [J. Discrete Algorithms 2013] runs in time $\Omega(n\sigma)$ on pseudoforests, we present a new algorithm 
based on dynamic range stabbing 
building the LCP array of $\Ginfsup$ in $O(n\log \sigma)$ time and $O(n\log\sigma)$ bits of working space. Combined with our reduction, we obtain the first efficient algorithm to build the LCP array of an arbitrary labeled graph. An implementation of our algorithm is publicly available at \url{https://github.com/regindex/Labeled-Graph-LCP}.
\end{abstract}

\newpage

\section{Introduction}\label{sec:intro}

The LCP array of a string --- storing the lengths of the longest common prefixes of lexicographic-adjacent string suffixes --- is a data structure introduced by Manber and Myers in \cite{MM93} that proved very useful in tasks such as speeding up pattern matching queries with suffix arrays \cite{MM93} and representing more compactly the suffix tree \cite{abouelhoda2004replacing}. As a more recent application of this array, Boucher \textit{et al.} augmented the BOSS representation of a de Bruijn graph with a generalization of the LCP array that supports the navigation of the underlying variable-order de Bruijn graph \cite{boucherdcc2015}. Recently, Conte \textit{et al.} \cite{conte2023dcc} and Cotumaccio \cite{NC-LCP-arxiv} extended this structure to Wheeler DFAs \cite{GagieMS:tcs17:wheeler} --- deterministic edge-labeled graphs admitting a total order of their nodes being compatible with the co-lexicographic order of the strings labeling source-to-node paths --- and, ultimately, arbitrary edge-labeled graphs. The main idea when generalizing the LCP array to a labeled graph, is to collect the lexicographic smallest $\inf_u$ and largest $\sup_u$ 
string entering each node $u$ (following edges backwards, starting from $u$), sorting them lexicographically, and computing the lengths of the longest common prefixes between lexicographically-adjacent strings in this sorted list \cite{conte2023dcc}. As shown by Conte \textit{et al.} \cite{conte2023dcc} and Cotumaccio \textit{et al.} \cite{cotumaccio2023spire,NC-LCP-arxiv}, such a data structure can be used to efficiently find Maximal Exact Matches (MEMs) on the graph's paths and to speed up the navigation of variable-order de Bruijn graphs.  

Importantly, \cite{conte2023dcc,cotumaccio2023spire,NC-LCP-arxiv} did not discuss efficient algorithms for building the LCP array of a labeled graph. The goal of our paper is to design such algorithms. 

\paragraph*{Overview of our contributions} 

Let $G$ be a directed labeled graph with $n$ nodes and $m$ edges labeled over an alphabet of cardinality $\sigma$. 
After introducing the main definitions and notation in Section \ref{sec:preliminaries}, in Section \ref{sec:pipeline} we describe our main three-steps pipeline for computing the LCP array of $G$: (1) \emph{Pre-processing}. Using the algorithms described in \cite{becker_et_al:LIPIcs.ESA.2023.15,cotumaccio:LIPIcs.ISAAC.2023.22,alanko2020soda}  we transform $G$ into a particular deterministic \emph{Wheeler pseudoforest} $\Ginfsup$ of size $O(n)$, that is, a deterministic Wheeler graph \cite{GagieMS:tcs17:wheeler} whose nodes have in-degree equal to 1.  
Graph $\Ginfsup$ compactly encodes the lexicographically- smallest and largest strings $\inf_u$ and $\sup_u$ (Definition \ref{def:inf sup}) leaving backwards (i.e. following reversed edges) every node $u$ of $G$, by means of the unique path entering the node. 
Using state-of-the-art algorithms, 
this step runs in $O(m)$ time and $O(m)$ words of space if $G$ is a Wheeler semi-DFA \cite{alanko2020soda},
and in $O(\min\{m+n^2,m\log n\})$ time and $O(m)$ words of space on arbitrary deterministic graphs  \cite{becker_et_al:LIPIcs.ESA.2023.15,cotumaccio:LIPIcs.ISAAC.2023.22}.
(2) \emph{LCP computation}: we describe a new compact-space algorithm computing the LCP array of $\Ginfsup$ (see below for more details); (3) \emph{Post-processing}: we turn the LCP array of $\Ginfsup$ into the LCP array of the original graph $G$, in $O(m)$ time and $O(m)$ words of space.

As far as step (2) is concerned, 
we turn our attention to
the algorithm of Beller \textit{et al.} \cite{beller2013computing}, working  on strings in $O(n\log\sigma)$ time and $O(n\log\sigma)$ bits of working space on top of the LCP array (which can be streamed to the output in the form of pairs $(i,LCP[i])$ sorted by their second component). The idea of this algorithm is to keep a queue of suffix array intervals, initially filled with the interval $[1,n]$. After extracting from the queue an interval $[i,j]$ corresponding to all suffixes prefixed by some string $\alpha$, the algorithm retrieves all \emph{distinct} characters $c_1,\dots, c_k$ in the Burrows-Wheeler transform \cite{burrows1994block} interval $BWT[i,j]$ and, by means of backward searching \cite{FM2000}, retrieves the intervals of strings $c_1\cdot \alpha, \dots, c_k\cdot \alpha$, writing an LCP entry at the beginning of each of those intervals (unless that LCP entry was not already filled, in which case the procedure does not recurse on the corresponding interval). Recently, Alanko \textit{et al.}~\cite{DBLP:conf/spire/AlankoBP23} generalized this algorithm to the BWT of the infimum sub-graph of a de Bruijn graph (also known as the SBWT \cite{DBLP:conf/acda/AlankoPV23}). 
In Section \ref{sec:stabbing} we first show a pseudoforest on which the natural generalization of Beller \textit{et al.}'s  algorithm \cite{beller2013computing} performs $\Omega(n\sigma)$ steps of forward search; this implies that, as a function of $\sigma$, this algorithm is \emph{exponentially} slower on graphs than it is on strings.
Motivated by this fact, we revisit the algorithm. 
For simplicity, in this paragraph, we sketch the algorithm on strings; see Section \ref{sec:stabbing} for the generalization to graphs.
Rather than working with a queue of suffix array intervals, we maintain a queue of LCP indexes $i\in [1,n]$. We furthermore keep a dynamic range-stabbing data structure on all the open intervals $(l,r]$ such that $BWT[l]=BWT[r]=c$ for some character $c$, and no other occurrence of $c$ appears in $BWT[l,r]$. When processing position $i$ with $LCP[i]=\ell$, we remove the intervals $(l,r]$ stabbed by $i$, and use them to derive new LCP entries of value $\ell+1$ (and new positions to be inserted in the queue) by backward-stepping from $BWT[l]$ and $BWT[r]$. 
On deterministic Wheeler pseudoforests, our algorithm runs in $O(n\log\sigma)$ time and uses $O(n\log\sigma)$ bits of working space (the LCP array can be streamed to output as in the case of Beller \textit{et al.} \cite{beller2013computing}).

Putting everything together (preprocessing, LCP of $\Ginfsup$, and post-processing), we prove (see Section \ref{sec:preliminaries} for all definitions):

\begin{theorem}\label{thm:main}
    Given a labeled graph $G$ with $n$ nodes and $m$ edges labeled over alphabet $[\sigma]$, with $\sigma \leq m^{O(1)}$, we can compute the LCP array of $G$ 
    in $O(m)$ words of space and $O(n\log\sigma + \min\{m\log n, m + n^2\})$ time. If $G$ is a Wheeler semi-DFA, the running time reduces to $O(n\log\sigma+m)$.
\end{theorem}

If the input graph $G$ is a Wheeler pseudoforest with all strings $\inf_u$ being distinct, represented compactly as an FM-index of a Wheeler graph \cite{GagieMS:tcs17:wheeler}, we can do even better: our algorithm terminates in $O(n\log\sigma)$ time while using just $O(n\log\sigma)$ bits of working space (Lemma \ref{lem:pseudoforest LCP}). 

We implemented our algorithm computing the LCP array of $G_{is}$ (Algorithm \ref{alg:lcpstab}) and made it publicly available at \url{https://github.com/regindex/Labeled-Graph-LCP}.

\section{Preliminaries}\label{sec:preliminaries}

We work with directed edge-labeled graphs $G=(V,E)$ on a fixed ordered alphabet $\Sigma = [\sigma] = \{1,\dots, \sigma\}$, where $E \subseteq V\times V\times \Sigma$
and, without loss of generality, $V = [n]$ for some integer $n>0$. 
Symbol $n=|V|$ denotes therefore the number of nodes of $G$. With $m=|E|$ we denote the number of edges of $G$. 
Without loss of generality, we assume that there are no nodes with both in-degree and out-degree equal to zero (such nodes can easily be treated separately in the problem we consider in this paper). In particular, this implies that $n  \in O(m)$.
We require that the alphabet's size is polynomial in the input's size: $\sigma \leq m^{O(1)}$.
Notation $\IN{u}$ and $\OUT{u}$, for $u\in V$, indicates the in-degree and out-degree of $u$, respectively. 
We say that $G$ is \emph{deterministic} if $(u, v', a), (u, v'', b) \in E \Rightarrow a\neq b$ whenever $v'\neq v''$. 
We say that 
any node $u\in V$ with $\IN{u} = 0$ is a \emph{source}, 
that $G$ is a \emph{semi-DFA} if $G$ is deterministic, has exactly one source, and all nodes are reachable from the source, 
and that $G$ is a \emph{pseudoforest} if and only if $\IN{u}= 1$ for all $u\in V$.
If $G=(V,E)$ is a pseudoforest, $\lambda(u)\in \Sigma$, for $u\in V$, denotes the character labeling the unique edge entering $v$.
We say that two labeled graphs $G=(V,E),G'=(V',E')$ on the same alphabet $\Sigma$ are \emph{isomorphic} if and only if there exists a bijection $\phi:V\rightarrow V'$ 
preserving edges and labels: for every 
$u,v\in V, u',v'\in V', a\in \Sigma$, 
$(u,v,a)\in E$ if and only if 
$(\phi(u),\phi(v),a)\in E'$.

If $\alpha = c_1c_2\cdots c_n \in\Sigma^*$ is a finite string, the notation $\overleftarrow\alpha = c_n c_{n-1}\cdots c_1$ indicates $\alpha$ reversed.
The notation $\Sigma^\omega$ indicates the set of \emph{omega strings}, that is, right-infinite strings of the form $c_1c_2c_3\dots$, with $c_i\in\Sigma$ for all $i\in \mathbb N^{>0}$.
As usual, $\Sigma^*$ and $\Sigma^+$ denote the sets of finite (possibly empty) strings and the set of nonempty finite strings from $\Sigma$, respectively.
For $\alpha = c_1c_2\dots \in\Sigma^\omega \cup \Sigma^*$,  $\alpha[i\dots]$ indicates the suffix $c_ic_{i+1}\dots$ of $\alpha$.
If $\alpha,\beta\in \Sigma^\omega \cup \Sigma^*$, we write $\alpha\prec\beta$ to indicate that $\alpha$ is lexicographically smaller than $\beta$ (similarly for $\preceq$: lexicographically smaller than or equal to). 
Symbol $\epsilon$ denotes the empty string, and it holds $\epsilon \prec \alpha$ for all $\alpha,\beta\in \Sigma^\omega \cup \Sigma^+$.
Given a set $S \subseteq \Sigma^\omega \cup \Sigma^*$ and a string $\alpha \in \Sigma^\omega \cup \Sigma^*$, notation $S \prec \alpha$ indicates $(\forall \beta\in S)(\beta \prec \alpha)$ (similarly for $\alpha \prec S$, $S \preceq \alpha$, and $\alpha \preceq S$).

If $G=(V,E)$ is deterministic, given $u\in V$ we denote with  $\mathsf{OUTL}(u)$ the sorted string of characters labeling outgoing edges from $u$: $\mathsf{OUTL}(u) = c_1\dots c_k$ if and only if $(\forall j \in [k])(\exists v\in V)\big((u,v,c_j)\in E\big)$, with $c_1 \prec c_2 \dots \prec c_k$. We write $c\in \mathsf{OUTL}(u)$, with $c\in \Sigma$, as a shorthand for $c \in \{\mathsf{OUTL}(u)[1], \dots, \mathsf{OUTL}(u)[k]\}$. Since we assume that $G$ is deterministic when we use this notation, it holds $\OUT{u} = |\mathsf{OUTL}(u)|$.

\emph{Wheeler graphs} were introduced by Gagie \textit{et al.} \cite{GagieMS:tcs17:wheeler}:

\begin{definition}\label{def:wheeler}
Let $ G = (V, E) $ be an edge-labeled graph, and let $ < $ be a strict total order on $ V $. 
For every $u,v\in V$, let  $u\le v$ indicate $u<v \vee u=v$. We say that $ < $ is a \emph{Wheeler order} for $G$ if and only if:
\begin{enumerate}
	\item (Axiom 1) For every $u,v\in V$, if $\IN{u}=0$ and $\IN{v}>0$ then $u < v$. 
    \item (Axiom 2) For every $ (u', u, a), (v', v, b) \in E $, if $ u < v $, then $ a \preceq b $.
    \item (Axiom 3) For every $ (u', u, a), (v', v, a) \in E $, if $ u < v $, then $ u' \le v' $.
\end{enumerate}
A graph $ G = (V, E) $ is Wheeler if it admits at least one Wheeler order.
\end{definition}





Let $u,v\in V$, $\alpha\in \Sigma^+$, and $c\in \Sigma$.
We write $u \stackrel{\alpha}{\rightsquigarrow} v$ to indicate that there exists a path from $u$ to $v$ labeled with string $\alpha$ (that is, we can go from $ u $ to $ v $ by following edges whose labels, when concatenated, yield $ \alpha $), and write $u \stackrel{c}{\rightarrow} v$ as an abbreviation for $(u,v,c) \in E$.

We use the symbol $I_u$ to denote the set of strings obtained starting from node $u$ and following edges backwards.  This process either stops at a node with in-degree 0 (thereby producing a finite string), or continues indefinitely (thereby producing an omega-string). Notice that this notation differs from \cite{alanko2020soda}, where edges are followed forwards; we use this slight variation since, as seen below, it is a more natural way to define the LCP array of a graph.
More formally:

\begin{definition}\label{def:I_q}
Let $G=(V,E)$ be a labeled graph. 
For $u\in V$, $I_u^\omega \subseteq \Sigma^\omega$ denotes the set of \emph{omega-strings} leaving backwards node $u$: 
$$
I_u^\omega =\{ c_1c_2\dots\ \in \Sigma^\omega\  :\ (\exists v_1,v_2, \dots \in V)(v_1\stackrel{c_1}{\rightarrow} u \wedge (\forall i \geq 2 )(v_i\stackrel{c_i}{\rightarrow} v_{i-1}) )\}.
$$
The symbol $I_u \subseteq \Sigma^\omega \cup \Sigma^*$ denotes instead the set of \emph{all} strings leaving backwards $u$:
$$
I_u =  I_u^\omega \cup \{ \overleftarrow \alpha \in \Sigma^+\  :\ (\exists v\in V)(\IN{v}=0 \wedge v \stackrel{\alpha}{\rightsquigarrow} u)\};
$$
If $\IN{u}=0$, we define $I_u=\{\epsilon\}$. 
\end{definition}


\begin{definition}[Infimum and supremum strings\cite{kim2023cpm}]\label{def:inf sup}
Let $G=(V,E)$ and $u\in V$. The infimum string $\inf_u^G = \inf I_u$ and the supremum string $\sup_u^G = \sup I_u$ relative to $G$ are defined as: 
\begin{align*}
\mathrm{inf}_u^G &= \gamma\in\Sigma^*\cup\Sigma^\omega \mbox{ s.t. } (\forall \beta\in\Sigma^*\cup\Sigma^\omega)(\beta 
\preceq I_u \rightarrow \beta \preceq \gamma \preceq I_u) \\
\mathrm{sup}_u^G &= \gamma\in\Sigma^*\cup\Sigma^\omega \mbox{ s.t. } (\forall \beta\in\Sigma^*\cup\Sigma^\omega)(I_u \preceq \beta \rightarrow 
I_u \preceq \gamma \preceq \beta) 
\end{align*}
When $G$ will be clear from the context, we will drop the superscript and simply write $\inf_u$ and $\sup_u$.
\end{definition}

Cotumaccio \cite{NC-LCP-arxiv} defines the LCP of a labeled graph as follows:

\begin{definition}\label{def:lcp}
    Let $ G = (V, E) $ be a labeled graph. Let $ \gamma_1 \preceq \gamma_2 \preceq \dots \preceq \gamma_{2n} $ be the 
    lexicographically-sorted strings $\inf_u$ and $\sup_u$, for all $u\in V$.
    The \emph{LCP array} $\LCP_G[2, 2n]$ of $ G $ is defined as $ \LCP_G [i] = \lcp (\gamma_{i - 1}, \gamma_i) $, where $ \lcp(\gamma_{i - 1}, \gamma_i) $ is the length of the longest common prefix between $ \gamma_{i - 1} $ and $ \gamma_i $. 
\end{definition}

In the above definition, observe that $ \LCP_G[i] = \infty $ if and only if $ \gamma_{i - 1} = \gamma_i $ and $\gamma_{i - 1}, \gamma_i \in \Sigma^\omega$. We are interested in this definition since, as shown in \cite{NC-LCP-arxiv}, it allows computing matching statistics on arbitrary labeled graphs.

If $G=(V,E)$ is a pseudoforest, then $I_u$ is a singleton for every $u\in V$ and the above definition can be simplified since $\inf_u=\sup_u$ for every $u\in V$. In this paper we will consider the particular case of pseudoforests for which all the $\inf_u(=\sup_u)$ are distinct. In this particular case, we define the \emph{reduced LCP array} as follows:

\begin{definition}\label{def:reduced lcp}
    Let $ G = (V, E) $ be a labeled pseudoforest such that $\inf_u\neq \inf_v$ for all $u\neq v \in V$. Let $ \gamma_1^* \prec \gamma_2^* \prec \dots \prec \gamma_n^* $ be the 
    lexicographically-sorted strings $\inf_u$, for all $u\in V$.
    The \emph{reduced LCP array} $\LCPx_G = \LCPx_G[2, n] $ of $ G $ is defined as $\LCPx_G [i] = \lcp (\gamma_{i - 1}^*, \gamma_i^*) $.
\end{definition}

Since pseudoforests will play an important role in our algorithms, we proceed by proving a useful property of the reduced LCP array of a pseudoforest.

\begin{lemma}\label{lem:downbyone}
	Let $ G = (V, E) $ be a labeled pseudoforest such that $\inf_u\neq \inf_v$ for all $u\neq v \in V$. Let $ 1 \le k < \infty $ be such that $ \LCPx_G[i] = k $ for some $ 2 \le i \le n $. Then, there exists $ 2 \le i' \le n $ such that $ \LCPx_G[i'] = k - 1 $.
\end{lemma}

\begin{proof}
Let $ \gamma_1^* \prec \gamma_2^* \prec \dots \prec \gamma_n^* $ be the lexicographically-sorted strings $\inf_u$, for all $u\in V$. For all $i\in [n]$ let $1\leq p(i) \leq n$ be the unique integer such that $\gamma_{p(i)}^* = \gamma_i^*[2\dots]$. Such integer always exists and it is unique, since the $\gamma_j^*$'s are all the strings leaving each node of $G$ (following edges backwards).
    
Since $ 1 \le k < \infty $, then $ \gamma_{i - 1}^*[1] = \gamma_i^*[1] $ and $ \gamma_{p(i - 1)}^* = \gamma_{i - 1}^*[2\dots] \prec \gamma_i^*[2\dots] = \gamma_{p(i)}^* $, which implies $ p(i - 1) < p(i)$. As a consequence, 
$ k = \LCPx_G[i] = \lcp(\gamma_{i - 1}^*, \gamma_i^*) = 1 + \lcp(\gamma_{p(i - 1)}^*, \gamma_{p(i)}^*) = 1 + \min_{p(i - 1) + 1 \le t \le p(i)} \lcp(\gamma_{t - 1}^*, \gamma_t^*) = 1 + \min_{p(i - 1) + 1 \le t \le p(i)} \LCPx_G[t] $, so there exists $ p(i - 1) + 1 \le i' \le p(i) $ such that $ k = 1 + \LCPx_G[i'] $, or equivalently, $ \LCPx_G[i'] = k - 1 $.
\end{proof}

\section{The pipeline: computing $\LCP_G$ from $G$}\label{sec:pipeline}

As mentioned in Section \ref{sec:intro}, we reduce the computation of $\LCP_G$ to three steps: (1) a pre-processing phase building a  Wheeler pseudoforest $\Ginfsup$, (2) the computation of the reduced LCP array of ${\Ginfsup}$, and (3) a post-processing phase yielding $\LCP_G$.
Steps (1) and (3) mainly use existing results from the literature and we illustrate them in this section. Our main contribution is step (2) for which in Section \ref{sec:stabbing} we describe a new algorithm.

\subsection{Pre-processing}\label{sec:preprocessing}

Let $G = (V,E)$ be the input graph.
As the first step of our pre-processing phase, we 
augment $\Sigma \leftarrow \Sigma \cup \{\#\}$ with a new symbol $\# (=0)$ lexicographically-smaller than all symbols in the original alphabet $\{1,\dots,\sigma\}$, and 
add a self-loop $u \stackrel{\#}{\rightarrow} u$ to all nodes $u\in V$ such that $\IN{u}=0$.
This will simplify our subsequent steps as now all strings leaving (backwards) any node belong to $\Sigma^\omega$; from now on, we will therefore work with graphs with no sources.
 
Consider the set $IS = \{\inf_u, \sup_u :\ u\in V\}$, and let $N=|IS| \leq 2n$. Note that $N$ could be strictly smaller than $2n$ since some nodes may share the same infimum/supremum string. 
Moreover, by the definition of infima and suprema strings, for each $\alpha \in IS$ it holds that $\alpha[2\dots] \in IS$: this is true because, if $\alpha$ is the infimum of $I_u$, then $u$ has a predecessor $v$ such that (i) $v \stackrel{\alpha[1]}{\rightarrow}u$ and (ii) $\alpha[2\dots]$ is the infimum of $I_v$ (the same holds for suprema strings). Let us give the following definition. 

\begin{definition}\label{def:Gis'}
    Given $G=(V,E)$ with no sources, let $IS = \{\inf_u, \sup_u :\ u\in V\}$. We denote with $\Ginfsup' = (IS, E_{is}')$ the labeled graph with edge set $E_{is}' = \{(\alpha[2\dots],\alpha,\alpha[1])\ :\ \alpha \in IS\}$.
\end{definition}

Observe that (i) each node $\alpha \in IS$ of $\Ginfsup'$ has exactly one incoming edge, and (ii) $\Ginfsup'$ is deterministic since $\alpha \stackrel{a}{\rightarrow} \beta$ and $\alpha \stackrel{a}{\rightarrow} \beta'$ imply $\beta = a\cdot \alpha = \beta'$. In other words, $\Ginfsup'$ is a deterministic pseudo-forest. 
In addition, $\inf_u \neq \inf_v$ for every $u \neq v \in IS$.
Let us prove that $\Ginfsup'$ is a Wheeler graph.

\begin{lemma}\label{lem:wheelercondition}
    The lexicographic order $\prec$ on the nodes of $\Ginfsup'$ is a Wheeler order.
\end{lemma}
\begin{proof}
    We prove that $\prec$ satisfies the three axioms of Definition \ref{def:wheeler}.

    (Axiom 1). This axiom holds trivially, since $\Ginfsup'$ has no nodes with in-degree 0.

    (Axiom 2). Let $ (\alpha', \alpha, \alpha[1]), (\beta', \beta, \beta[1]) \in E_{is}'$, with $ \alpha \prec \beta $. Then, by definition of the lexicographic order $\prec$, it holds $\alpha[1] \preceq \beta[1]$. 
    
    (Axiom 3). Let $ (\alpha[2\dots], \alpha, a), (\beta[2\dots], \beta, a) \in E_{is}'$, with $a=\alpha[1]=\beta[1]$ and $ \alpha \prec \beta$. Then, by the definition of $\prec$ it holds $\alpha[2\dots] \prec \beta[2\dots]$.
\end{proof}

Importantly, we remark that our subsequent algorithms will only require the topology and edge labels of $\Ginfsup'$ to work correctly. In other words, any graph \emph{isomorphic} to $\Ginfsup'$ will work, and it will not be needed to compute \emph{explicitly} the set $IS$ (an impractical task, since $IS$ contains omega-strings). 
Figure \ref{fig:graphs} shows an example of labeled graph $G$ (with sources) pre-processed to remove sources and converted to such a graph isomorphic to $\Ginfsup'$. 
We can compute such a graph using recent results in the literature:

\begin{theorem}\label{thm:compute G_is}
    Let $G = (V,E)$ be a labeled graph with no sources and alphabet of size $\sigma \leq m^{O(1)}$.
    Let moreover $V_i = \{u_i\ :\ u\in V\}$ and 
    $V_s = \{u_s\ :\ u\in V\}$ be two duplicates of $V$.
    Then, we can compute a graph $\Ginfsup = ([N],E_{is})$ being isomorphic to $\Ginfsup' = (IS, E_{is}')$ (Definition \ref{def:Gis'}), together with a function $\mathsf{map} : V_i \cup V_s \rightarrow [N]$  such that:
    \begin{enumerate}
        \item for every $u\in V$, $\inf_u^G = \inf_{\mathsf{map}(u_i)}^{\Ginfsup}$ and $\sup_u^G = \sup_{\mathsf{map}(u_s)}^{\Ginfsup}$, and
        \item the total order $<$ on the integers coincides with the Wheeler order of $\Ginfsup$. In particular, for all $i,j \in [N]$,  $i<j$ if and only if $\inf_i^{\Ginfsup} = \sup_i^{\Ginfsup} \prec \inf_j^{\Ginfsup} = \sup_j^{\Ginfsup}$. 
    \end{enumerate}
    Function $\mathsf{map}$ is returned as an array of $2n=2|V|$ words, so that it can be evaluated in constant time. 
    $\Ginfsup$ and $\mathsf{map}$ can be computed from $G$ in 
    $O(m+n^2)$ time \cite{cotumaccio:LIPIcs.ISAAC.2023.22} or in $O(m\log n)$ time \cite{becker_et_al:LIPIcs.ESA.2023.15}. If $G$ is a Wheeler semi-DFA, the running time reduces to $O(m)$  \cite{alanko2020soda}. All these algorithms use $O(m)$ words of working space.
\end{theorem}

In Appendix \ref{sec:convert to Ginfsup} we discuss how Theorem \ref{thm:compute G_is} can be obtained using the results in \cite{cotumaccio:LIPIcs.ISAAC.2023.22,becker_et_al:LIPIcs.ESA.2023.15,alanko2020soda} (which were originally delivered for a different purpose: computing the maximum co-lex order \cite{cotumacciojacm2023, cotumacciosoda2021, cotumacciodcc2021, cotumaccio2024stacs}).
Figure \ref{fig:graphs} shows an example of $\Ginfsup$ (right) for a particular labeled graph $G$ (left).
Figure \ref{fig:LCPstar} (right) shows the nodes $[N]$ of such a graph $\Ginfsup$, together with the strings entering in each node, sorted lexicographically, and their longest common prefix array $\LCPx_{\Ginfsup}$. In Figure \ref{fig:LCPstar} (left) we sort the duplicated nodes ($V_i\cup V_s$) of $G$ by their infima ($V_i$) and suprema ($V_s$) strings, and show the mapping $\mathsf{map} : V_i \cup V_s \rightarrow [N]$ in the second and third columns.

\begin{figure}[!th]
\centering
    \resizebox{\textwidth}{!}{
\begin{tikzpicture}[>=Stealth, node distance = 1.4cm, on grid, auto]
    \tikzset{
        mynode/.style={circle, draw, minimum size=0.8cm},
        myarrow/.style={->, >=Stealth},
        scale=0.5
    }
    
    \node[mynode] (1) {1};
    \node[mynode, right=of 1] (2) {2};
    \node[mynode, below=of 2] (3) {3};
    \node[mynode, left=of 3] (4) {4};
    \node[mynode, right=of 2] (5) {5};
    \node[mynode, below=of 5] (6) {6};
    \node[mynode, below=of 6] (7) {7};
    \node[mynode, below=of 7] (8) {8};
    \node[mynode, right=of 8] (9) {9};
    \node[mynode, left=of 7] (10) {10};
    \node[mynode, right=of 5] (11) {11};
    \node[mynode, right=of 11] (12) {12};
    \node[mynode, below=of 12] (13) {13};
    \node[mynode, below=of 13] (14) {14};
    \node[mynode, right=of 14] (15) {15};
    \node[mynode, left=of 14] (16) {16};

    \draw[myarrow] (1) edge node {A} (2);
    \draw[myarrow] (2) edge node {T} (3);
    \draw[myarrow] (3) edge node {A} (4);
    \draw[myarrow] (4) edge node {T} (1);
    \draw[myarrow] (5) edge node {A} (6);
    \draw[myarrow] (6) edge[left] node {C} (7);
    \draw[myarrow] (7) edge node {T} (8);
    \draw[myarrow] (8) edge node {A} (9);
    \draw[myarrow] (10) edge node {A} (7);
    \draw[myarrow] (11) edge node {C} (12);
    \draw[myarrow] (12) edge node {C} (13);
    \draw[myarrow] (13) edge node {T} (14);
    \draw[myarrow] (14) edge node {A} (9);
    \draw[myarrow] (14) edge[loop below] node {T} (14);
    \draw[myarrow] (14) edge node {A} (15);
    \draw[myarrow] (13) edge node {T} (16);
    \draw[myarrow] (9) edge[right, near end] node {C} (6);
    \draw[myarrow] (12) edge node {T} (6);

    \draw (16.5,2) -- (16.5,-10); 

    %
    %
    %
    %

    \node[mynode, right=4cm of 12] (11_is) {11};
    \node[mynode, right=of 11_is] (4_is) {4};
    \node[mynode, below=of 11_is] (15_is) {15};
    \node[mynode, right=of 15_is] (6_is) {6};
    \node[mynode, right=of 4_is] (1_is) {1};
    \node[mynode, below=of 1_is] (2_is) {2};
    \node[mynode, below=of 2_is] (10_is) {10};       
    \node[mynode, left=of 10_is] (3_is) {3};
    \node[mynode, right=of 1_is] (7_is) {7};
    \node[mynode, below=of 7_is] (8_is) {8};
    \node[mynode, below=of 8_is] (13_is) {13};
    \node[mynode, right=of 13_is] (5_is) {5};
    \node[mynode, right=of 7_is] (12_is) {12};
    \node[mynode, below=of 12_is] (9_is) {9};
    \node[mynode, right=of 9_is] (14_is) {14};

    \draw[myarrow] (11_is) edge[bend left] node {A} (4_is);
    \draw[myarrow] (4_is) edge[bend left] node {T} (11_is);
    \draw[myarrow] (15_is) edge[loop below] node {T} (15_is);
    \draw[myarrow] (15_is) edge node {A} (6_is);
    
    \draw[myarrow] (1_is) edge[loop above] node {\#} (1_is);
    \draw[myarrow] (1_is) edge node {A} (2_is);    
    \draw[myarrow] (2_is) edge node {T} (10_is);    
    \draw[myarrow] (10_is) edge node {A} (3_is);
    
    \draw[myarrow] (1_is) edge node {C} (7_is);
    \draw[myarrow] (7_is) edge node {C} (8_is);
    \draw[myarrow] (8_is) edge node {T} (13_is);
    \draw[myarrow] (13_is) edge node {A} (5_is);
    
    \draw[myarrow] (7_is) edge node {T} (12_is);
    \draw[myarrow] (12_is) edge node {C} (9_is);
    \draw[myarrow] (9_is) edge node {T} (14_is);

    

\end{tikzpicture}
}
    \caption{\emph{Left}: a labeled graph~$G$. \emph{Right}: a graph $\Ginfsup$ isomorphic to $\Ginfsup'$ (Definition \ref{def:Gis'}) satisfying Theorem~\ref{thm:compute G_is}. Note that in $\Ginfsup$ the node numbering coincides with the Wheeler order.
    } \label{fig:graphs}
\end{figure}
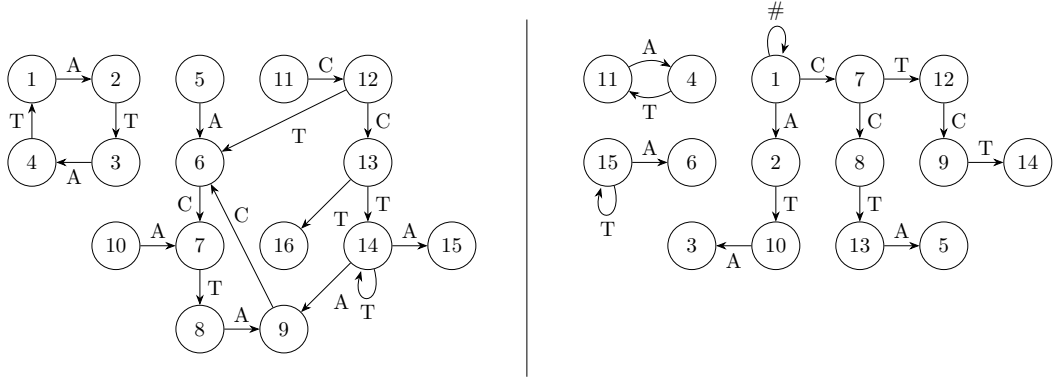

\begin{figure}[!tbh]
    \centering
        \resizebox{\textwidth}{!}{
\begin{tabular}{|ccclc|ccclc|c|cll|}
\cline{1-10} \cline{12-14}
$j$ & $[N]$ & \multicolumn{1}{c}{$V_i \cup V_s$} & \multicolumn{1}{c}{$\gamma_j$} & $\mathsf{LCP}_G$ & $j$ & $[N]$ & \multicolumn{1}{c}{$V_i \cup V_s$} & \multicolumn{1}{c}{$\gamma_j$} & $\mathsf{LCP}_G$ && $[N]$ & \multicolumn{1}{c}{$\gamma_j^*$} & $\LCPx_{\Ginfsup}$ \\
\cline{1-10} \cline{12-14}
1 & 1 & $10_i$ & $\varepsilon$ & - & 17 & 7 & $12_i$ & {\tt C} & 0 && 1 & {\tt \#\#\#\#\#...} & - \\
2 & 1 & $10_s$ & $\varepsilon$ & 0 & 18 & 7 & $12_s$ & {\tt C} & 1 && 2 & {\tt A\#\#\#\#...} & 0 \\
3 & 1 & $11_i$ & $\varepsilon$ & 0 & 19 & 8 & $13_i$ & {\tt CC} & 1 && 3 & {\tt ATA\#\#...} & 1 \\
4 & 1 & $11_s$ & $\varepsilon$ & 0 & 20 & 8 & $13_s$ & {\tt CC} & 2 && 4 & {\tt ATATA...} & 3 \\
5 & 1 & $5_i$ & $\varepsilon$ & 0 & 21 & 9 & $7_s$ & {\tt CTC} & 1 && 5 & {\tt ATCC\#...} & 2 \\
6 & 1 & $5_s$ & $\varepsilon$ & 0 & 22 & 10 & $8_i$ & {\tt TA} & 0 && 6 & {\tt ATTTT...} & 2 \\
7 & 2 & $6_i$ & {\tt A} & 0 & 23 & 11 & $1_i$ & {\tt TATAT...} & 2 && 7 & {\tt C\#\#\#\#...} & 0 \\
8 & 2 & $7_i$ & {\tt A} & 1 & 24 & 11 & $1_s$ & {\tt TATAT...} & $\infty$ && 8 & {\tt CC\#\#\#...} & 1 \\
9 & 3 & $9_i$ & {\tt ATA} & 1 & 25 & 11 & $3_i$ & {\tt TATAT...} & $\infty$ && 9 & {\tt CTC\#\#...} & 1 \\
10 & 4 & $2_i$ & {\tt ATATA...} & 3 & 26 & 11 & $3_s$ & {\tt TATAT...} & $\infty$ && 10 & {\tt TA\#\#\#...} & 0 \\
11 & 4 & $2_s$ & {\tt ATATA...} & $\infty$ & 27 & 12 & $6_s$ & {\tt TC} & 1 && 11 & {\tt TATAT...} & 2 \\
12 & 4 & $4_i$ & {\tt ATATA...} & $\infty$ & 28 & 13 & $14_i$ & {\tt TCC} & 2 && 12 & {\tt TC\#\#\#...} & 1 \\
13 & 4 & $4_s$ & {\tt ATATA...} & $\infty$ & 29 &  13 & $16_i$ & {\tt TCC} & 3 && 13 & {\tt TCC\#\#...} & 2 \\
14 & 5 & $15_i$ & {\tt ATCC} & 2 & 30 & 13 & $16_s$ & {\tt TCC} & 3 && 14 & {\tt TCTC\#...} & 2 \\
15 & 6 & $15_s$ & {\tt ATTTT...} & 2 & 31 & 14 & $8_s$ & {\tt TCTC} & 2 && 15 & {\tt TTTTT...} & 1 \\
16 & 6 & $9_s$ & {\tt ATTTT...} & $\infty$ & 32 & 15 & $14_s$ & {\tt TTTTT...} & 1 & &  &  &  \\
\cline{1-10} \cline{12-14}
\end{tabular}
} 
    \caption{\emph{Left:} lexicographically-sorted infima and suprema strings $\gamma_j$ of graph $G=(V,E)$ of Figure~\ref{fig:graphs}, along with the nodes of $V$ they reach (subscripted using the duplicates $V_i$ and $V_s$ of $V$ to show whether the string is an infimum or a supremum), and array $\LCP_G$. 
    The second ($[N]$) and third ($V_i\cup V_s$) columns of the table show the mapping  $\mathsf{map} : V_i \cup V_s \rightarrow [N]$.
    \emph{Right:} The sorted infima (equivalently, suprema) $\gamma_j^*$ of graph $\Ginfsup$ of Figure~\ref{fig:graphs}, and the array $\LCPx_{\Ginfsup}$.}
    \label{fig:LCPstar}
\end{figure}

\subsection{Post-processing}\label{sec:postprocessing}

In Section \ref{sec:preprocessing} we have shown how to convert the input $G=(V,E)$ into a Wheeler pseudoforest $\Ginfsup = ([N],E_{is})$ encoding the infima and suprema strings of $G$. In Section \ref{sec:stabbing} we will show how to compute the reduced $\LCPx_{\Ginfsup}$ of $\Ginfsup$. Here, we discuss the last step of our pipeline, converting $\LCPx_{\Ginfsup}$ into $\LCP_G$.

Let $\gamma_1^* \prec \gamma_2^* \prec \dots \prec \gamma_N^*$ be the sorted strings $\inf_j$, for $j\in [N]$, relative to graph $\Ginfsup$, and $\gamma_1 \preceq \gamma_2 \preceq \dots \preceq \gamma_{2n}$ be the sorted strings $\inf_u, \sup_u$, for $u\in V$, relative to graph $G$.
By construction (Section \ref{sec:preprocessing}), the former sequence of strings is obtained from the latter by performing these two operations: (1) duplicates are removed, and (2) finite strings $\gamma_i$ are turned into omega-strings of the form $\gamma_i\cdot \#^\omega$ (see Figure \ref{fig:LCPstar}, right table).
This means that $\LCP_G$ is almost the same as $\LCPx_{\Ginfsup}$, except in maximal intervals $\LCP_G[i+1,j]$ such that $\gamma_i = \gamma_{i+1} = \dots = \gamma_j$. In those intervals, we have $\LCP_G[i+1,j] = |\gamma_i|$; notice that this value could be either finite (if $\gamma_i\in \Sigma^*$) or infinite (if $\gamma_i\in \Sigma^\omega$). 

As an example, consider the interval 
$\LCP_G[10,13] = (3,\infty,\infty,\infty)$ in Figure \ref{fig:LCPstar} (left). This interval corresponds to strings \texttt{ATATA...}, and corresponds to $\LCPx_{\Ginfsup}[4]=3$ (right). The first value $\LCP_G[10]$ is equal to $\LCPx_{\Ginfsup}[4]=3$, while the others, $\LCP_G[11,13] = (\infty,\infty,\infty)$ are equal to the length ($\infty$) of the omega-string \texttt{ATATA...}. A similar example not involving an omega string is $\LCP_G[7,8] = (0,1)$ (string \texttt{A}), corresponding to $\LCPx_{\Ginfsup}[2]=0$.

Given $\LCPx_{\Ginfsup}$, $\Ginfsup$, and $\mathsf{map}$ (see Theorem \ref{thm:compute G_is}) it is immediate to derive $\LCP_G$ in $O(m)$ worst-case time and $O(m)$ words of working space, as follows.
First of all, we sort $V_i \cup V_s$ (the two duplicates of $V$, see Theorem \ref{thm:compute G_is}) according to the order given by the integers $\mathsf{map}(x)$, for $x\in V_i \cup V_s$.
Let $u^1, u^2, \dots, u^{2n}$
be the corresponding sequence of sorted nodes, i.e. such that $\mathsf{map}(u^1) \le \mathsf{map}(u^2) \le \dots \le \mathsf{map}(u^{2n})$.
The second step is to identify 
the above-mentioned maximal intervals $\LCP_G[i+1,j]$: these are precisely the maximal intervals such that $\mathsf{map}(u^i) = \mathsf{map}(u^{i+1}) = \dots = \mathsf{map}(u^{j})$. For each such interval, we set $\LCP_G[i] = \LCPx_{\Ginfsup}[\mathsf{map}(u^i)]$ and $\LCP_G[i+1] = \LCP_G[i+2] = \dots \LCP_G[j] = |\gamma_i|$. In order to compute the length $|\gamma_i|$, observe that $\gamma_i = \gamma^*_{\mathsf{map}(u^i)}$ if $\gamma_i \in \Sigma^\omega$, and $\gamma_i\cdot \#^\omega = \gamma^*_{\mathsf{map}(u^i)}$ if $\gamma_i \in \Sigma^*$. Then, a simple DFS visit of $\Ginfsup$ starting from nodes $u$ with a self-loop $u \stackrel{\#}{\rightarrow} u$ reveals whether we are in the former or latter case, and allows computing the length (DFS depth) of $\gamma_i$ in the latter.

\section{Computing the LCP array of $\Ginfsup$}\label{sec:stabbing}

In view of Sections \ref{sec:preprocessing} and \ref{sec:postprocessing}, the core task to solve in order to compute $\LCP_G$ is the computation of $\LCPx_{\Ginfsup}$. 
To make notation lighter, in this section we denote the input graph with symbol $G=(V,E)$ and assume it is a deterministic Wheeler pseudoforest such that $\inf_u\neq \inf_v$ for all $u\neq v\in V$. 
In the rest of the section, $n$ and $m$ denote the number of nodes and edges of $G$.
The goal of our algorithms is to compute $\LCPx_G$.

\begin{figure}[h!]
\centering
\scalebox{0.7}{
\centering
\begin{tikzpicture}[->,>=stealth', semithick, auto, scale=1]
\tikzset{every state/.style={minimum size=30pt}}
\node[state] (v0) at (0,0) {$u$};

\node[state] (v1) at (2,0) {$v_1$};
\node[state] (v2) at (4,0) {$v_2$};
\node (vc) at (6,0) {$\cdots$};
\node[state] (vn1) at (8,0) {$v_{n-1}$};
\node[state] (vn) at (10,0) {$v_{n}$};

\node[state] (z1) at (13,2) {$z_{1}$};
\node[state] (z2) at (13,0.5) {$z_{2}$};
\node (zc) at (13,-0.75) {$\cdots$};
\node[state] (zn) at (13,-2) {$z_{n}$};

\draw (v0) edge [loop left] node [] {$ a $} (v0);
\draw (v0) edge [above] node [] {$ b $} (v1);
\draw (v1) edge [above] node [] {$ b $} (v2);
\draw (v2) edge [above] node [] {$ b $} (vc);
\draw (vc) edge [above] node [] {$ b $} (vn1);
\draw (vn1) edge [above] node [] {$ b $} (vn);
\draw (vn) edge [above] node [] {$ c_1 $} (z1);
\draw (vn) edge [above] node [] {$ c_2 $} (z2);
\draw (vn) edge [dashed] node [] {} (zc);
\draw (vn) edge [above] node [] {$ c_{n} $} (zn);

\end{tikzpicture}
}
\caption{A Wheeler pseudoforest with $V=\{u < v_1 < \cdots < v_n < z_1 < \cdots < z_n\}$ and Wheeler order $<$
where the number of forward search steps in the natural generalization of Beller \textit{et al.}'s algorithm is $\Theta(n\sigma)$. 
The range of nodes (BWT interval) reached by a path suffixed by $b^i$, for $1\le i\le n$, is $[v_i,v_n]$. 
The algorithm right-extends (via \emph{forward search}) each of these ranges by all the $n$ characters $c_1, \dots, c_n$; when extending with $c_i$, we obtain the unary range $[z_i]$ of nodes reached by a path suffixed by $b^ic_i$. This means that Beller \textit{et al.}'s algorithm will perform in total $\Theta(n\sigma) = \Theta(n^2)$ forward search steps. Intuitively, our solution to this problem will be to extend \emph{just one} of those ranges by  $c_1, \dots, c_n$. We achieve this by resorting to a dynamic range stabbing data structure. }\label{fig:counterxamplelinear2}
\end{figure}
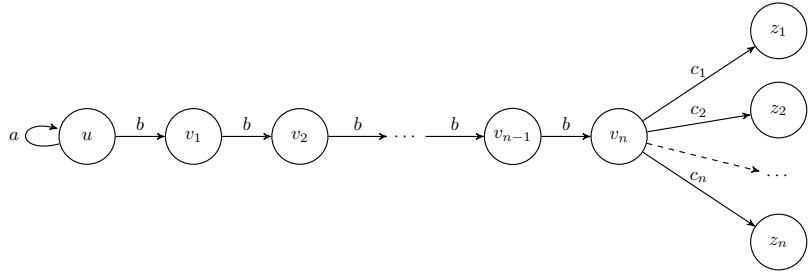

To solve this problem, we first focused our attention on the algorithm of Beller \textit{et al.} \cite{beller2013computing}, computing the LCP array from the Burrows-Wheeler Transform \cite{burrows1994block} of the input string in $(1+o(1))\cdot n\log\sigma + O(n)$ bits of working space (including the indexed BWT and excluding the LCP array, which however can be streamed to output in order of increasing LCP values) and $O(n\log \sigma)$ time. 
However, as we briefly show in Figure \ref{fig:counterxamplelinear2}, we realized that the natural generalization of this algorithm to pseudoforests runs in $\Omega(n\sigma)$ time in the worst case. 
Motivated by this fact, in this section we re-design the algorithm by resorting to the dynamic interval stabbing problem (e.g., see \cite{yakov11:intervalstab}), achieving running time $O(n\log \sigma)$ on deterministic Wheeler pseudoforests. This will require designing a novel dynamic range-stabbing data structure that could be of independent interest.

Let $v_1 < v_2 < \dots < v_n$ denote the Wheeler order of $G$.
We define:


\begin{definition}[bridge]\label{def:bridge}
    For $1\le l<r\le n$ and $c\in\Sigma$, a triplet $(l,r,c)$ is said to be a \emph{bridge} of $G$ if and only if (i) both $v_l$ and $v_r$ have an outgoing edge labeled with $c$ and (ii) for every $k$ such that $l<k<r$, $v_k$ does not have any outgoing edge labeled with $c$.
\end{definition}

Consider two consecutive nodes (in the Wheeler order) $v_{i-1}$ and $v_{i}$ with the same incoming label $c=\lambda(v_{i-1})=\lambda(v_i)$ for some $1<i\le n$. On deterministic pseudoforests there is a one-to-one correspondence between the set of such node pairs and the set of bridges: 

\begin{lemma}\label{lem: bridge-LCP one-to-one}
    The following bijection exists between the set of bridges and the set of pairs $(v_{i-1},v_i)$ such that $\lambda(v_{i-1})=\lambda(v_{i})$. 
    Let $\lambda(v_{i-1})=\lambda(v_{i})=c$, and let $v_{l}$ and $v_{r}$ be the nodes such that $v_l \stackrel{c}{\rightarrow} v_{i-1}$ and $v_r \stackrel{c}{\rightarrow} v_{i}$. Then, $(l,r,c)$ is a bridge. Conversely, 
    for every bridge $(l,r,c)$, letting $ v_h $ and $ v_i $ be the unique nodes such that $v_l \stackrel{c}{\rightarrow} v_{h} $ and $v_r \stackrel{c}{\rightarrow} v_{i} $, we have $ h = i - 1 $.
\end{lemma}
\begin{proof}

Given such nodes $v_{i-1}$ and $v_{i}$, $v_{l}$ and $v_{r}$ are unambiguously determined because every node has exactly one incoming edge. To see that $(l,r,\lambda(v_i))$ is a bridge, notice that for every edge $(u,u',\lambda(v_i))$, if $ u' < v_{i - 1} $, then by Axiom 3 and determinism $ u < v_l $, and if $ v_i < u' $, then similarly $ v_r < u $, so for every $ k $ such that $ l < k < r $, $ v_k $ does not have an outgoing edge labeled $ \lambda (v_i) $. For the reverse implication, notice that $ h \not = i $ because every node has exactly one incoming edge, and it cannot be $ i < h $ otherwise from Axiom 3 we would obtain $ v_r < v_l $. Hence $ h \le i - 1 $. If we had $ h < i - 1 $, then by Axiom 2 the unique edge entering $ v_{i - 1} $ should be labeled $ c $, and from Axiom 3 and determinism its start node $ k $ should satisfy $ l < k < r $, so $ (l, r, c) $ would not be a bridge.
\end{proof}


The intuition behind our algorithm is as follows.
Given a bridge $(l,r,c)$, suppose that $\lcp(\gamma^*_{l},\gamma^*_{r})= d \ge 0$. Let $v_{i-1}$ and $v_{i}$ be the nodes such that $v_l \stackrel{c}{\rightarrow} v_{i-1}$ and $v_r \stackrel{c}{\rightarrow} v_{i}$ (see Lemma \ref{lem: bridge-LCP one-to-one}). 
These nodes can be obtained from $v_l$ and $v_r$ by one forward search step.
Then, we have $\gamma^*_{i-1}=c\gamma^*_{l}$ and $\gamma^*_{i}=c\gamma^*_{r}$, thus $\lcp(\gamma^*_{i-1},\gamma^*_{i})=\lcp(\gamma^*_{l},\gamma^*_{r})+1=d+1$. 
By Lemma~\ref{lem:wheelercondition}, the Wheeler order $v_1 < \dots < v_n$ corresponds to the lexicographic order of the node's incoming strings $\gamma^*_1 \prec \dots \prec \gamma^*_n$, hence $\lcp(\gamma^*_l,\gamma^*_r)=\min_{j\in(l,r]}\lcp(\gamma^*_{j-1},\gamma^*_{j})=\min_{j\in(l,r]}\LCPx_G[j]$.
Therefore, the minimum value $d = \min_{j\in(l,r]}\LCPx_{G}[j]$ within the left-open interval $(l,r]$ corresponding to a bridge $(l,r,c)$ yields $\LCPx_{G}[i]=\lcp(\gamma^*_{i-1},\gamma^*_{i}) = d+1$. 
This observation stands at the core of our algorithm: if we compute LCP values in nondecreasing order, then the position $j_{min} = \mathrm{argmin}_{j\in(l,r]}\LCPx_{G}[j]$ ($1< j_{min} \le n$) of the first (smallest) generated LCP value inside $(l,r]$ ``stabs'' interval $(l,r]$.
This yields $\LCPx_G[i] = \LCPx_G[j_{min}]+1$. 
After this interval stabbing query, bridge $(l,r,c)$ is removed from the set of bridges since the resulting LCP value $\LCPx_G[i]$ has been correctly computed once for all.

Our procedure for computing $\LCPx_{G}$ is formalized in Algorithm~\ref{alg:lcpstab}.
The algorithm takes as input a deterministic Wheeler pseudoforest $G$ represented as an FM index (Lemma \ref{lem:FMWG} below) and outputs all pairs $(i,\LCPx_G[i])$, one by one in a streaming fashion, in nondecreasing order of their second component (i.e. LCP value). This is useful in space-efficient applications where one cannot afford storing the whole LCP array in $n\log n$ bits, see for example \cite{prezza2021space}.


\begin{lemma}\label{lem:FMWG}
    A deterministic Wheeler pseudoforest $G$ can be represented succinctly in $(1+o(1))\cdot n\log\sigma + O(n)$ bits of space with a data structure (FM index of a Wheeler graph \cite{GagieMS:tcs17:wheeler}) supporting the following queries in $O(\log\sigma)$ time: 
    \begin{enumerate}
        \item $G.\mathsf{forward\_step}(i,c)$: given $1< i\le n$ and a character $c\in\Sigma$, let $k\geq i$ be the smallest integer such that $v_k$ has an outgoing edge labeled with $c$. This query returns the integer $i'$ such that $v_k \stackrel{c}{\rightarrow} v_{i'}$, or $\bot$ if such $k$ does not exist.
        \item $\mathsf{OUTL}(v_i)[j]$: given a node $v_i$ and an index $j\in[\sigma]$, return the $j$-th outgoing label of $v_i$. Return $\bot$ if  $j>\OUT{v_i}$.
        \item $\lambda(v_i)$: given $i\in [n]$, return the incoming label of $v_i$. 
    \end{enumerate}
\end{lemma}
\begin{proof}
    Following \cite{GagieMS:tcs17:wheeler}, 
    we represent $G$ as a triple
    $(\mathsf{C}, \mathsf{OUT},\mathsf{L}) \in \{0,\dots,n\}^\sigma \times \{0,1\}^{2n}\times [\sigma]^n$ defined as:
    \begin{itemize}
        \item $\mathsf{OUT} = 0^{\OUT{v_1}}1\cdot 0^{\OUT{v_2}}1 \cdots 0^{\OUT{v_n}}1$ is the concatenation of the outdegrees of nodes $v_1, \dots, v_n$, written in unary,  
        \item $\mathsf{L} = \mathsf{OUTL}(v_1)\cdot \mathsf{OUTL}(v_2)\cdots\mathsf{OUTL}(v_n)$ is the concatenation of all the labels of the node's outgoing edges in Wheeler order, and 
        \item $\mathsf{C}[c] = |\{u\in V\ :\ \lambda(u)\prec c\}|$, $c\in \Sigma$, denotes the number of nodes whose incoming edge is labelled with a character $c'$ such that $c'\prec c$. Importantly, $\mathsf{C}$ is not actually stored explicitly;   as we show below, any $\mathsf{C}[c]$ can be retrieved from 
        $\mathsf{L}$ in $O(\log\sigma)$ time. 
    \end{itemize}
    The only difference with \cite{GagieMS:tcs17:wheeler} (where arbitrary Wheeler graphs are considered) is that a pseudoforest has $n$ nodes and $n$ edges, and all nodes have in-degree equal to 1. This simplifies the structure, since we do not need to store in-degrees. 
     $\mathsf{L}$ is encoded with a wavelet tree \cite{GrossiGV03:wavelettree} and $\mathsf{OUT}$ with a succinct bitvector data structure \cite{RamanRS07:rrrvector}. A root-to leaf traversal of the wavelet tree of $\mathsf{L}$ is sufficient to retrieve any  $\mathsf{C}[c]$ in $O(\log\sigma)$ time at no additional space cost  (see for example \cite[Alg 3]{Schnattinger13bidirectional}).
    This representation uses $n\log\sigma + O(n)$ bits of space and supports the following operations in $O(\log\sigma)$ time: (1) random access to any of the arrays $\mathsf{C}, \mathsf{OUT},\mathsf{L}$, (2)  $\mathsf{L}.\mathsf{rank}_c(j)$, returning the number of occurrences of $c$ in  $\mathsf{L}[1,j]$, and (3) $\mathsf{OUT}.\mathsf{select}_1(j)$, returning the position of the $j$-th occurrence of bit $1$ in bitvector $\mathsf{OUT}$. 
    
    Using these queries, we can solve query (1) as follows: $G.\mathsf{forward\_step}(i,c) = \mathsf{C}[c]+\mathsf{L}.\mathsf{rank}_c(\mathsf{OUT}.\mathsf{select}_1(i-1)-(i-1))+1$. If $\mathsf{L}.\mathsf{rank}_c(\mathsf{OUT}.\mathsf{select}_1(i-1)-(i-1))+1$ exceeds the number of characters equal to $c$ in $\mathsf{L}$ (we discover this in $O(\log\sigma)$ time using the wavelet tree on $\mathsf{L}$), the query returns $\bot$.
    Query (2) is answered as follows: $\mathsf{OUTL}(v_i)[j] = \mathsf{L}[\mathsf{OUT}.\mathsf{select}_1(i-1)-(i-1)+j]$. If $j$ exceeds the out-degree of $v_i$ (we discover this in constant time using rank and select queries on $\mathsf{OUT}$), the query returns $\bot$. Query (3) $\lambda(v_i)$ can be solved with a root-to-leaf visit of the wavelet tree of $\mathsf{L}$, in $O(\log\sigma)$ time (range quantile queries, see \cite{gagie09rangequantile}). 
\end{proof}

\begin{algorithm}[h!]
\caption{Given a Wheeler pseudoforest $G$, compute $\LCPx_G$. In Line \ref{alg:removed}, $ {\mathcal I}.~\mathsf{stab\_and\_remove}(i) $ stabs and removes bridges from $ \mathcal{I} $.}\label{alg:lcpstab}
\begin{algorithmic}[1]
    \State $ \LCPx_G \gets \text{ Array $ \LCPx_G [2, n] $, with values initialized to $\infty$} $ 
    \State $W \gets \emptyset$\Comment{$W$: queue of integer pairs of the form $(i,\LCPx_G[i])$}
    \label{alg:interval init queue}
    \State $\mathcal I \gets \{(l,r,c)\in[n]\times[n]\times\Sigma\ :\ \text{$(l,r,c)$ is a bridge of $G$} \}$\Comment{$\mathcal I$: bridges of $G$}\label{alg:build bridges}
    \label{alg:interval line:c-bridge intervals}
    \State \textbf{for all} $1< i\le n$ \textbf{such that} $\lambda(v_{i-1})\prec\lambda(v_{i})$: $W.\mathsf{push}(i, 0)$
    \label{alg:lnterval line:initborder}
    
    \While{$ W \neq \emptyset$}\label{alg:while}
        \State $(i, d) \gets W.\mathsf{pop}()$  \label{alg:pop}
        \State \textbf{output} $(i,d)$ \Comment{Stream pair $(i,\LCPx_G[i])$ to output}\label{alg:output}
        \State $R\gets {\mathcal I}.~\mathsf{stab\_and\_remove}(i)$
        \label{alg:lnterval line:stab}\Comment{$R \subseteq \Sigma $: set of labels of bridges stabbed and removed}\label{alg:removed}
        \For{$c \in R $} \label{alg:for}
            \State $i' \gets G.\mathsf{forward\_step}(i,c)$ \label{alg:interval line:fwd}
            \State $W.\mathsf{push}(i', d+1)$\label{alg:push2}
        \EndFor
    \EndWhile
\end{algorithmic}
\end{algorithm}

We proceed by commenting the pseudocode and proving its correctness and complexity.
In Line \ref{alg:interval line:c-bridge intervals} of Algorithm~\ref{alg:lcpstab}, we compute the set $\mathcal{I}$ of bridges of the input graph using Lemma \ref{lem: interval stabbing}. 
Each bridge $(l,r,c)$ will ``survive'' in $\mathcal I$ until any $\LCPx_G[i]$ with $i\in (l,r]$ is computed. 
Set $\mathcal{I}$ is represented as a dynamic range stabbing data structure (Lemma \ref{lem: interval stabbing} below) on the set of \emph{character-labeled} intervals $\{(l,r]_c\ :\ (l,r,c)\ \mathrm{is\ a\ bridge}\}$, where notation $(l,r]_c$ indicates a left-open interval labeled with character $c$. We require this data structure to support interval stabbing and interval deletion queries. General solutions for this problem solving both queries in amortized $O(\log n/\log\log n)$ time exist \cite{yakov11:intervalstab}. 
While in the general case this is optimal, in our particular case observe that, by Definition \ref{def:bridge}, no more than $\sigma$ intervals get stabbed by a given $i \in [n]$.
We exploit this property to develop a more efficient (if $\log\sigma = o(\log n/ \log\log n)$) dynamic range stabbing data structure (for the full proof, see Appendix \ref{ref:proof range stabbing}):

\begin{lemma}\label{lem: interval stabbing}
    Given a Wheeler pseudoforest $G$ represented with the data structure of Lemma \ref{lem:FMWG}, in $O(n\log\sigma)$ time and $O(n\log\sigma)$ bits of working space we can build a \emph{dynamic interval stabbing} data structure $\mathcal{I}$ of $O(n\log\sigma)$ bits 
    representing the set of bridges of $G$ (Definition \ref{def:bridge}) and    
    answering the following query: $\mathcal{I}.\mathsf{stab\_and\_remove}(i)$, where $i\in[n]$. Letting $S = \{(l,r,c)\in \mathcal I\ :\ l<i\le r\}$ be the set of stabbed bridges, this query performs the following two operations:
    \begin{enumerate}
        \item it returns the set of characters $R = \{c\ :\ (l,r,c)\in S\}$ labeling bridges stabbed by $i$, and 
        \item it removes those bridges: $\mathcal I \leftarrow \mathcal I \setminus S$.
    \end{enumerate}
    Let $\ell = \sum_{(l,r,c)\in S} (r-l+1)$ be the total length of the stabbed bridges. 
    The query is solved in $O(\log \sigma + |R| + \ell/\sigma)$ time.
\end{lemma}
\begin{proof}
   (Sketch, see Appendix \ref{ref:proof range stabbing} for all the details). We divide the nodes $v_1, \dots, v_n$ into non-overlapping blocks $v_{k\sigma},\dots, v_{(k+1)\sigma}$ of $\sigma$ nodes each, for $k=0,\dots, n/\sigma - 1$ (assume $\sigma$ divides $n$ for simplicity). 
   Let $I$ be the set of labeled intervals $(l,r]_c$ corresponding to all the bridges $(l,r,c)$ of $G$; the bridges of $G$ can be reconstructed in $O(\log\sigma)$ time each using operation $\mathsf{OUTL}(v_i)[j]$ of Lemma \ref{lem:FMWG}.
   Each interval $(l,r]_c\in I$
   overlapping $t>1$ blocks (i.e. $l+1 \leq k\sigma \leq (k+t-2)\sigma  < r$ for some $k\in\{0,\dots, n/\sigma-1\}$) is broken into $t$ ``pieces'' $(l_1=l,r_1]_c, \dots, (l_t,r_t=r]_c$: a suffix of a block, followed by full blocks, followed by a prefix of a block. 
   Each piece $(l_i,r_i]_c$, overlapping the $k$-th block for some $k$, is inserted in interval set $I_k$. The pieces  $(l_1,r_1]_c, \dots, (l_t,r_t]_c$ are connected using a doubly-linked list.
   Intervals of $I$ fully contained in a block are not split in any piece and just inserted in $I_k$, with $k$ being the block they overlap. 
   Since no more than $\sigma$ intervals can pairwise intersect at any point $i\in[n]$, for every $k$ at most $\sigma$ ``pieces'' of at most $\sigma$ intervals are inserted in $I_k$: in total, the 
    number of intervals in all the sets $I_k$ is therefore $\sum_{k=0}^{n/\sigma-1}|I_k| \in O(|I| + \sigma \cdot n/\sigma) = O(n)$ (because $|I|\in O(n)$ by Lemma \ref{lem: bridge-LCP one-to-one}).
   We build an interval tree data structure $\mathcal T(I_k)$ (\cite[Ch 8.8]{book:computational:geometry}, \cite[Ch 17.3]{book:introduction:algorithm}) on each $I_k$. Interval stabbing queries are answered locally (on the tree associated with the block containing the stabbing position $i$). Interval deletion queries require to also follow the linked list associated with the deleted interval, to delete all $\ell/\sigma$ ``pieces'' of the original interval (of length $\ell$) of $I$. 
   Each interval piece is deleted in constant time since we do not need to re-balance the tree.
   Our claim follows by observing that $|I_k| \in O(\sigma^2)$ for every $k$ (because no more than $\sigma$ intervals can pairwise intersect at any point), so (i) each $\mathcal T(I_k)$ is built in $O(|I_k|\log|I_k|) = O(|I_k|\log\sigma)$ time (overall, all trees are built in $O(n\log\sigma)$ time), and (ii) each pointer (tree edges and linked list pointers) uses just $O(\log\sigma)$ bits:  observe that linked list pointers always connect intervals belonging to adjacent trees $\mathcal T(I_k)$, $\mathcal T(I_{k\pm 1})$, so they point inside a memory region of size $O(|I_k|)\subseteq O(\sigma^2)$ and thus require $O(\log |I_k|) \subseteq O(\log\sigma)$ bits each.
\end{proof}

After building data structure $\mathcal I$, in Line \ref{alg:lnterval line:initborder} we identify all integers $1<i\le n$ such that $\LCPx_G[i]=0$: these correspond to consecutive nodes $v_{i-1}, v_i$ with different incoming labels: $\lambda(v_{i-1}) \neq \lambda(v_i)$. 

We keep the following invariant before and after the execution of each iteration of the \texttt{while} loop at Line \ref{alg:while}: for every $1<i\le n$, \emph{exactly one} of the following three conditions holds. (i) the pair $(i,\LCPx_G[i])$ has already been output at line \ref{alg:output}, (ii) $(i,d) \in W$ for some $d\ge 0$, in which case it holds that $\LCPx_G[i] = d$ or (iii) $(l,r,c) \in \mathcal I$, where $(l,r,c)$ is the bridge associated with $v_i$ according to Lemma~\ref{lem: bridge-LCP one-to-one}. 

The invariant clearly holds before entering the \texttt{while} loop, because for every $1<i\le n$ we either push $(i,0)$ in $W$ at Line \ref{alg:lnterval line:initborder} whenever $\lambda(v_{i-1}) \neq \lambda(v_{i})$ (thereby satisfying condition (ii) since $\LCPx_G[i] = \lcp(\gamma^*_{i-1}, \gamma^*_{i})=0$), or insert $(l,r,c)$ in $\mathcal{I}$ at Line \ref{alg:build bridges}, where $(l,r,c)$ is the bridge associated with $v_i$ (thereby satisfying condition (iii)). 
At this point, condition (i) does not hold for any $1<i\le n$.
Notice that by Definition \ref{def:bridge}, no bridge is associated with $v_i$ such that $\lambda(v_{i-1}) \neq \lambda(v_{i})$ (and vice versa), so the invariant's conditions are indeed mutually exclusive.

We show that the invariant holds after every operation in the body of the main loop.
Assume we pop $(i,d)$ from $W$ at Line \ref{alg:pop}. Then, this means that (before popping) condition (ii) of our invariant holds, and in particular $\LCPx_G[i] = d$. At line \ref{alg:output} the algorithm correctly outputs $(i,d=\LCPx_G[i])$, so condition (i) of the invariant now holds for position $i$ (while condition (ii) does not hold anymore, and condition (iii) did not hold even before: remember that the three conditions are mutually exclusive). The invariant is not modified (so it still holds) for the other positions $i'\neq i$.

In Line \ref{alg:lnterval line:stab}, we retrieve and remove every bridge $(l,r,c)$ such that $l<i\le r$. Let $v_{i'}$ be the node associated with bridge $(l,r,c)$ (Lemma \ref{lem: bridge-LCP one-to-one}). The fact that we remove $(l,r,c)$ from $\mathcal I$ temporarily invalidates the invariant for $i'$ (none of (i-iii) hold), but in the \texttt{for} loop at Line \ref{alg:for} we immediately re-establish the invariant by pushing in $W$ pair $(i',d+1)$ and observing that indeed $\LCPx_G[i']=d+1$ (i.e. condition (ii) of the invariant holds for position $i'$). To see that $\LCPx_G[i']=d+1$ holds true first observe that, since (i) we use a queue $W$ for pairs $(i,\LCPx_G[i])$, 
(ii) initially (Line \ref{alg:lnterval line:initborder}), we only push in $W$ pairs of the form $(i,0)$, and 
(iii) whenever we pop a pair $(i,d)$ we push pairs of the form $(j,d+1)$ for some $i,j\in [n]$, LCP values are popped in line \ref{alg:pop} in nondecreasing order. In particular, for every $d\ge 0$, no pair of the form $(i,d+1)$ is popped from the queue until all pairs of the form $(j,d)$ are popped. From this observation and since $i$ is the first integer stabbing bridge $(l,r,c)$ in Line \ref{alg:lnterval line:stab} (since we remove bridges from $\mathcal I$ immediately after they are stabbed), it must be $\lcp(\gamma^*_{l}, \gamma^*_{r}) = \min_{l<j\le r}\LCPx_G[j]= \LCPx_G[i] = d$. Then, since $\gamma^*_{i'-1} = c\gamma^*_{l}$ and $\gamma^*_{i'} = c\gamma^*_{r}$ hold, we have that $\LCPx_G[i'] = \lcp(\gamma^*_{i'-1}, \gamma^*_{i'}) = d+1$.

We proved that our invariant always holds true; in particular, it holds when the algorithm terminates. 
The fact that conditions (i-iii) are mutually exclusive, immediately implies that no LCP value is output more than once, i.e. the first components of the output pairs $(i,\LCPx_G[i])$ are all distinct.

At the end of the algorithm's execution, $W=\emptyset$ holds.
To prove that the algorithm computes every LCP value, suppose for a contradiction, that there exists $i\in[n]$ such that $(i,\LCPx_G[i])$ is never output in Line \ref{alg:output}. Without loss of generality, choose $i$ yielding the smallest such $\LCPx_G[i]$. 
Note that $\LCPx_G[i]>0$, since all pairs $(j,\LCPx_G[j]=0)$ are inserted in $W$ at Line \ref{alg:lnterval line:initborder}, thus they are output at Line \ref{alg:output}.
Then, condition (i) of our invariant does not hold for $i$. Also condition (ii) cannot hold, otherwise it would be $(i,\LCPx_G[i])\in W \neq \emptyset$. We conclude that condition (iii) must hold for $i$: $(l,r,c) \in \mathcal I$, where $(l,r,c)$ is the bridge associated with $v_i$ by Lemma~\ref{lem: bridge-LCP one-to-one}. In turn, this implies that no pair $(j,\LCPx_G[j])$ for $l<j\le r$ has been output in Line \ref{alg:output} (otherwise, such a $j$ stabbing $(l,r,c)$ would have caused the removal of $(l,r,c)$ from $\mathcal I$). By Lemma \ref{lem: bridge-LCP one-to-one}, $v_{l} \stackrel{c}{\rightarrow} v_{i-1}$ and $v_{r} \stackrel{c}{\rightarrow} v_{i}$ hold.
Since we assume that $\inf_u\neq \inf_v$ for all $u\neq v \in V$, we can apply Lemma \ref{lem:downbyone} and obtain that it must hold $ \LCPx_G[j]=\LCPx_G[i]-1$ for some $l<j\le r$, which contradicts to minimality of $\LCPx_G[i]$. 
We conclude that the algorithm computes every LCP value.

Next, we analyze the algorithm's running time and working space. By Lemma \ref{lem: interval stabbing}, $\mathcal I$ is built in $O(n\log\sigma)$ time using $O(n\log\sigma)$ bits of working space.
The \texttt{while} loop at Line \ref{alg:while} iterates at most $O(n)$ times because (i) at most $n$ elements are pushed into the queue $W$ at the beginning (Line \ref{alg:lnterval line:initborder}), and (ii) an element $(i,d)\in\mathbb{N}^2$ can be pushed into the queue at Line \ref{alg:push2} only after a bridge is stabbed and removed; thus at most $|\mathcal I| \in O(n)$ elements can be pushed into the queue.
As a result, Line \ref{alg:lnterval line:stab} (query $\mathcal I.\mathsf{stab\_and\_remove}(i)$) is executed $O(n)$ times. Recall (Lemma \ref{lem: interval stabbing}) that such a query runs in $O(\log \sigma + |R| + \ell/\sigma)$ time, where $|R|$ is the number of characters labeling stabbed intervals (equivalently, the number of stabbed intervals since no two intervals labeled with the
same character can intersect) and $\ell$ is the total cumulative length of the stabbed intervals. Since overall the calls to
$\mathcal I.\mathsf{stab\_and\_remove}(i)$ will ultimately remove
all bridges from $\mathcal I$, we conclude that the sum of all cardinalities $|R|$ equals $|\mathcal I| \in O(n)$, and the sum of all  cumulative lengths $\ell$ equals the sum of all the bridges' 
lengths: $\sum_{(l,r,c)\in \mathcal I}(r-l+1) \in O(n\sigma)$ (because no two intervals labeled with the
same character can intersect). 
Appying Lemma \ref{lem: interval stabbing} we conclude that, overall, all calls to $\mathcal I.\mathsf{stab\_and\_remove}(i)$ cost $O(n\log\sigma + n + n\sigma / \sigma) =  O(n\log\sigma)$ time.
Line~\ref{alg:interval line:fwd} takes $O(\log\sigma)$ time by Lemma \ref{lem:FMWG}. 
We represent the queue $W$ in $O(n)$ bits of space, using the same strategy of Beller \textit{et al.} (see \cite{beller2013computing} for all details): $W$ is represented internally with two queues $W_{t}$ and $W_{t+1}$, containing pairs of the form $(i,t)$ and $(i,t+1)$, respectively (in fact, only the first element $i$ of the pair needs to be stored). Pairs are popped from the former queue and pushed into the second. As long as $|W_{t+1}| \leq n/\log n$, $W_{t+1}$ is represented as a simple vector of integers. As soon as $|W_{t+1}| > n/\log n$, we switch to a packed bitvector representation of $n$ bits ($n/\log n$ words) marking with a bit set all $i$ such that $(i,t+1)\in W_{t+1}$. When $W_{t}$ becomes empty, we delete it, create a new queue $W_{t+2}$, and start popping from $W_{t+1}$ and pushing into $W_{t+2}$. If $W_{t+1}$ is still represented as a vector of integers, popping is trivial. Otherwise (packed bitvector), all integers in $W_{t+1}$ can be popped in overall $O(n/\log n + |W_{t+1}|) \subseteq O(|W_{t+1}|)$ time using bitwise operations.

We finally obtain:

\begin{lemma}\label{lem:pseudoforest LCP}
    Given a deterministic Wheeler pseudoforest $G = (V,E)$ such that $\inf_u \neq \inf_v$ for all $u\neq v \in V$    
    represented with the data structure of Lemma \ref{lem:FMWG}, the reduced LCP array $\LCPx_G$ of $G$ can be computed in $O(n\log\sigma)$ time and $O(n\log\sigma)$ bits of working space on top of the input. The algorithm does not allocate memory for the output array $\LCPx_G$: 
    the entries of this array are streamed to output in the form of pairs $(i,\LCPx_G[i])$ sorted by their second component, from smallest to largest.
\end{lemma}

Putting everything together (pre-processing, Lemma \ref{lem:pseudoforest LCP}, and post-processing), we obtain the main result of our paper, Theorem \ref{thm:main}.

\bibliography{mybibliography}

\newpage

\appendix

\section{Proof of Theorem \ref{thm:compute G_is}}\label{sec:convert to Ginfsup}

Let us prove Theorem \ref{thm:compute G_is}. In general, $\mathsf{map}(\cdot)$ yields the lexicographical rank of infima and suprema strings of the nodes in the input graph $G$. Intuitively, each edge in $G_{is}$ captures how the infimum and supremum string of a node is formed in terms of the infimum and supremum string of another node. For the details, we give descriptions in two different cases: (i) a general case where we can sort the infima/suprema strings of a graph in $O(m\log n)$ time \cite{becker_et_al:LIPIcs.ESA.2023.15} or in $O(n^2+m)$ time \cite{cotumaccio:LIPIcs.ISAAC.2023.22}, and (ii) a special case where a Wheeler semi-DFA is given so that we can compute the Wheeler order in $O(m)$ time \cite{alanko2020soda}.

\subsection{General case}

The algorithms in \cite{becker_et_al:LIPIcs.ESA.2023.15} and \cite{cotumaccio:LIPIcs.ISAAC.2023.22} compute the lexicographical ranks (equivalent to $\mathsf{map}(\cdot)$ in Theorem~\ref{thm:compute G_is}) of the infima and suprema strings associated with the nodes. Note that the setting used in \cite{becker_et_al:LIPIcs.ESA.2023.15} and \cite{cotumaccio:LIPIcs.ISAAC.2023.22} is slightly different, because the purpose is different: computing the maximum co-lex order \cite{cotumacciojacm2023, cotumacciosoda2021, cotumacciodcc2021} of a DFA (which is indeed characterized by the infima and suprema strings associated with each node \cite{kim2023cpm}). However, we can still apply those algorithms to our setting without significant modifications. The papers \cite{becker_et_al:LIPIcs.ESA.2023.15} and \cite{cotumaccio:LIPIcs.ISAAC.2023.22} assume that the input graph is a DFA, but the algorithms more generally compute the lexicographic ranks of the infima and suprema of an arbitrary graph. Moreover, the papers assume that all incoming edges entering a single node have the same label (\emph{input consistency}); this is in not restrictive, because in our setting we can simply remove, for each node, all the incoming edges with non-minimum (non-maximum) label among all incoming edges, which does not affect the infima (the suprema).

Once we have computed the ranks, we can immediately build $G_{is}$ from $ G $ is $ O(m) $ time by (i) creating a node for each infimum and each supremum, (ii) adding exactly one incoming edge for each node, consistently with the infimum or the supremum corresponding to the node and (iii) merging nodes with the same rank.

\subsection{Wheeler semi-DFAs}

The algorithm in \cite{alanko2020soda} computes the Wheeler order of a semi-DFA $G=(V,E)$, if it exists. The difference from the previous case is that we need to compute $\mathsf{map}(\cdot)$ from the Wheeler order. For the relation between $\mathsf{map}(\cdot)$ and the Wheeler order, we have the following results from \cite{kim2023cpm}.

\begin{lemma}
    \label{lem: Wheeler vs infsup}
    Let $<$ be the Wheeler order of a semi-DFA $G=(V,E)$. Then for every $u,v\in V$ such that $u<v$ the following hold:
    \begin{enumerate}
    \item  $\inf^G_u\preceq \sup^G_u\preceq \inf^G_v$.
    \label{lem: Wheeler vs infsup: infsup order}
    
    \item $\inf^G_u=\sup^G_u$ holds if and only if there exists a unique path from the source to $u$. 
    \label{lem: Wheeler vs infsup: infu eq supu}
    
    \item If $\sup^G_u=\inf^G_v$, then $\sup^G_u=\inf^G_v$ is a string of infinite length.
    \label{lem: Wheeler vs infsup: supu eq infv}
    
    \end{enumerate}
\end{lemma}
\begin{proof}
    Point (1) is immediate from \cite[Theorem~10]{kim2023cpm} and the fact that $<$ is a  total order. Points (2) and (3) follow from \cite[Observation~8]{kim2023cpm}. 
\end{proof}

Let $v_1<v_2<\cdots<v_n$ be the Wheeler order of a semi-DFA. Recall that $\mathsf{map}(\cdot)$ is consistent with the lexicographical ranks of the infima and suprema strings. From Lemma~\ref{lem: Wheeler vs infsup}-(\ref{lem: Wheeler vs infsup: infsup order}), we immediately obtain $\mathsf{map}(v_{1,i})\le \mathsf{map}(v_{1,s}) \le \mathsf{map}(v_{2,i})\le \mathsf{map}(v_{2,s})\le \cdots \le \mathsf{map}(v_{n,i})\le \mathsf{map}(v_{n,s})$ where, for $1\le j\le n$, $v_{j,i}$ and $v_{j,s}$ are nodes corresponding to infima and suprema strings of $v_j$. Therefore, to compute $\mathsf{map}(\cdot)$ completely, we need to determine (i) if $\inf^G_{v_j}=\sup^G_{v_j}$ for $1\le j\le n$, and (ii) if $\sup^G_{v_j}=\inf^G_{v_{j+1}}$ for $1\le j< n$.

By Lemma~\ref{lem: Wheeler vs infsup}-(\ref{lem: Wheeler vs infsup: infu eq supu}), we can find $v_j$'s such that $\inf^G_{v_j}=\sup^G_{v_j}$ by a simple graph traversal from the source to check if there is more than one path reaching each node.

What is left to compute $\mathsf{map}(\cdot)$ is to find $1\le j<n$ such that $\sup^G_{v_j}=\inf^G_{v_{j+1}}$. Consider graphs $G_i$ (and $G_s$, resp.) that encodes the infima (and suprema, resp.) strings, which can be obtained simply by removing edges that do not contribute to infima (and suprema, resp.) strings. For example, we can construct $G_i$ as follows: for every node $v\in V$, we remove all its incoming edges but the edge from the smallest node (in the Wheeler order).
From Lemma~\ref{lem: Wheeler vs infsup}-(\ref{lem: Wheeler vs infsup: supu eq infv}), when $\sup^G_{v_j}=\inf^G_{v_{j+1}}$, the string of concern should be of infinite length; hence, the backward walk from $v_j$ (and $v_{j+1}$) on graph $G_s$ (and $G_i$) representing $\sup^G_{v_j}$ (and $\inf^G_{v_{j+1}}$, resp.) must end up with a cycle on each graph. Therefore we need to detect such a pair of cycles to determine if $\sup^G_{v_j}=\inf^G_{v_{j+1}}$.

Observe that each of $G_i$ and $G_s$ is a pseudoforest where each component has at most one cycle, and the other nodes are reachable from the cycle in the same component. Thus we can obtain the set of cycles in $G_i$ and $G_s$ by performing topological sorting on each graph with reversed edges. Note that there exists a cycle $u\stackrel{\gamma}{\rightsquigarrow}u$ with $u\in V$ in $G_i$ (in $G_s$, resp.) iff $\inf^G_u=\overleftarrow\gamma^\omega$ ($\sup^G_u=\overleftarrow\gamma^\omega$, resp.). To compare cycles in $G_i$ and $G_s$ efficiently, we pick the smallest node (in the Wheeler order) for each cycle. Let $C$ be the set of nodes representing each cycle; the intuition is that if such a pair of cycles exist, then their smallest nodes must be adjacent in the Wheeler order. Sort $C$ in the Wheeler order, then for every pair of consecutive nodes in $C$, we compare the strings labeling their corresponding cycles. Since every string labeling a cycle in a Wheeler graph must be aperiodic, we can compare the strings labeling each cycle using a number of steps bounded by the cycle lengths. Each corresponding pair of nodes in each detected pair of cycles will have the same $\mathsf{map}(\cdot)$ value. Since each cycle is checked at most twice, this can be done in linear time from a given Wheeler order. 

After identifying the cycles labeled with the same string, we can also detect the other nodes for which the supremum string of one node is the same as the infimum string of the other by (i) sorting the remaining nodes (i.e., those not consisting of cycles in $G_i$ and $G_s$) in each component by topological order with ties broken by the Wheeler order, and (ii) checking their incoming label and (iii) verifying if any consecutive nodes are connected from the nodes that have been identified to have the same $\mathsf{map}(\cdot)$ value. As a result, we can compute $\mathsf{map}(\cdot)$ in linear time given the Wheeler order of a semi-DFA.

\section{Proof of Lemma \ref{lem: interval stabbing}}\label{ref:proof range stabbing}

We prove Lemma \ref{lem: interval stabbing}:  given a Wheeler pseudoforest $G$ represented with the data structure of Lemma \ref{lem:FMWG}, in $O(n\log\sigma)$ time and $O(n\log\sigma)$ bits of working space we can build a \emph{dynamic interval stabbing} data structure $\mathcal{I}$ of $O(n\log\sigma)$ bits 
representing the set of bridges of $G$ (Definition \ref{def:bridge}) and    
answering the following query: $\mathcal{I}.\mathsf{stab\_and\_remove}(i)$, where $i\in[n]$. Letting $S = \{(l,r,c)\in \mathcal I\ :\ l<i\le r\}$ be the set of stabbed bridges, this query performs the following two operations:
\begin{enumerate}
    \item it returns the set of characters $R = \{c\ :\ (l,r,c)\in S\}$ labeling bridges stabbed by $i$, and 
    \item it removes those bridges: $\mathcal I \leftarrow \mathcal I \setminus S$.
\end{enumerate}
Let $\ell = \sum_{(l,r,c)\in S} (r-l+1)$ be the total length of the stabbed bridges. 
The query is solved in $O(\log \sigma + |R| + \ell/\sigma)$ time.

\begin{proof}
    An interval tree (cf. \cite[Ch 8.8]{book:computational:geometry}, \cite[Ch 17.3]{book:introduction:algorithm}) on universe $[\sigma]$ is a balanced binary tree built over an interval set $I\subseteq [\sigma]\times [\sigma]$ where each node $x$ is associated with a position $p(x)\in [\sigma]$. The position $p(u)\in[\sigma]$ of the root $u$ is chosen so as to divide approximately in half the number of intervals 
    not intersecting $p(u)$. The root $u$ stores all intervals intersecting $p(u)$ (i.e. $(l,r]$ such that $l<p(u)\le r$), duplicated in two doubly-linked lists: one sorted by the interval's starting positions and the other by ending position. The two copies of the same interval in the two lists are linked so as to enable fast deletions (read below). The intervals stored in the root are removed from $ I $. If $y$ and $z$ are the left and right children of $u$, respectively, it holds $p(y) < p(u) < p(z)$. The remaining intervals in $ I $ are recursively distributed to the left and right child. This structure layout allows easily to label intervals: we will build interval trees over labeled left-open intervals denoted as $(l,r]_c$ (where $c\in \Sigma$ is the interval's label).
    A stabbing query with query position $i$ starts from the root $u$; if $i<p(u)$ (the other case $ p(u) < i $ is symmetric, and the case $ i = p(u) $ is simpler because we only have to retrieve the intervals stored in $ u $), it scans the list of intervals overlapping $p(u)$ sorted by their \emph{starting} position until they do not overlap $i$. After all such intervals are retrieved, the procedure continues recursively in the left child of $u$.
    As a result, an interval tree on $N$ intervals over universe $[\sigma]$ uses space $O(N\log(\sigma+N))$ 
    bits and supports stabbing queries in $O(\log N+t)$ time, where $t$ is the number of output intervals. 
    An interval can be deleted in $O(1)$ time given its location (a pointer of $O(\log N)$ bits) in the tree, by simply removing the two interval's copies in the two linked lists containing the interval (this may create nodes whose associated linked lists are empty, but this is not an issue: the tree's topology is not modified after a deletion, so the queries' running times do not change).
    Given the intervals, the interval tree data structure can be constructed in $O(N\log N)$ time using $O(N\log( \sigma+N))$ bits of space.

    Let $I=\{(l,r]_c:(l,r,c)\in\mathcal{I}\}$ be the set of character-labelled open intervals corresponding to the bridges of $G$. 
    Lemma \ref{lem: bridge-LCP one-to-one} implies that $|I| \leq n$.
    If $\sigma\ge n$ our claim immediately follows by building a classic interval tree on $I$, thus we can assume $\sigma<n$ in the remainder of the proof.
    
    We divide the set $V=\{v_1 < v_2 < \cdots < v_n\}$ of nodes (where $<$ denotes the Wheeler order) into $n/\sigma$ non-overlapping blocks $V_k=\{v_{k\sigma+i}\in V:1\le i\le \sigma\}$ of size $\sigma$ each, for all $0\le k< n/\sigma$. Without loss of generality, we assume that $\sigma$ divides $n$. Otherwise, the last block contains less than $\sigma$ nodes and our arguments still hold.

    The next step is to break intervals of set $I$ according to how they overlap with each $V_k$. For each $k$, we define the set $I_k=\{(l,r]_c \cap (k\sigma,(k+1)\sigma]\ :\ (l,r]_c\in I\} \setminus \{\emptyset_c:c\in\Sigma\}$ of non-empty intersections of intervals in $I$ with the open interval $(k\sigma,(k+1)\sigma]$ corresponding to block $V_k$.
    In this definition, the intersection $(a,b]_c \cap (d,e]$ of a labeled interval with a non-labeled one yields the labeled interval $(f,g]_c$, where $(f,g] = (a,b] \cap (d,e]$.

    Intuitively, 
    we will build an interval tree $\mathcal T(I_k)$ on each $I_k$ and, using doubly-linked lists, connect intervals (belonging to adjacent trees $\mathcal T(I_k)$, $\mathcal T(I_{k+1})$) 
    corresponding to pieces of the the same interval (spanning multiple blocks) in the original set $I$.    
    Note that, by definition of bridge, no more than $\sigma$ intervals can pairwise intersect so each $I_k$ will contain at most $\sigma^2$ intervals. 
    It follows that a relative pointer to an interval inside $\mathcal T(I_k)$ will use $O(\log|I_k|) = O(\log\sigma)$ bits.
    Moreover, by construction each tree $\mathcal T(I_k)$ contains intervals contained in range $(k\sigma,(k+1)\sigma]$; we will therefore store \emph{relative} intervals inside the tree: rather than storing interval $(l,r]$, we will actually store interval $(l-k\sigma, r-k\sigma]$. In this way, both the intervals and tree/list pointers of $\mathcal T(I_k)$ use $O(\log\sigma)$ bits each (for clarity of exposition, we omit this offset $-k\sigma$ in what follows).

    We now show how to compute each $\mathcal T(I_k)$ sequentially for $k=0, \dots,  n/\sigma-1$, in $O(n\log\sigma)$ time and using $O(n\log\sigma)$ bits of space. To this end, let us show how to compute each $ I_k $ sequentially.

    For simplicity, during the algorithm we will add some dummy intervals to some $ I_k $'s. Moreover, some additional intervals will be marked as dummy at the end of the algorithm. Dummy intervals do not correspond to real bridges. We start by inserting a dummy interval $(-1,0]_c$ in $I_0$ for all $c\in\Sigma$.
    We keep a \emph{global} array $P[1,\sigma]$ for all $c\in \Sigma$, to be re-used with every $\mathcal T(I_k)$. 
    We will process nodes (and thus blocks) in Wheeler order: $v_1, \dots, v_n$.
    Symbol $v_i$ will denote the currently-processed node, and $k =  i/\sigma  - 1$ the block containing $v_i$ (i.e. $v_i\in V_k$). 
    When we start processing block $V_k$, entry $P[c]$ will contain a pointer to either a dummy interval or to the ``piece'' of an interval labeled with $c$ that continues from the previous block. 
    During construction, we update $P$ so that, when we start processing node $v_i$, 
    $P[c]$ is a pointer to the interval $(l,r]_c$ maximizing $r\le i$ (in the current or previous block). Initially ($i=1$, $k=0$), $P[c]$ points to $(-1,0]_c$ for each $c\in \Sigma$. Note that, since each $P[c]$ points to an interval either in the previous or current block, each pointer uses $O(\log\sigma)$ bits, therefore $P$ use $O(\sigma\log\sigma) \subseteq O(n\log\sigma)$ bits of space.

    Let us define the following operation, creating and inserting an interval labelled with $c(\in\Sigma)$ ending in node $v_i(\in V)$ into $\mathcal T(I_k)$: 

    \begin{definition}[$\mathsf{insert}(i,c)$]\label{def:insert interval}
        Suppose we have processed nodes $v_1, \dots, v_{i-1}$.
        Let $P[c]$ point to the previous $c$-labeled interval $(l',r']_c$ encountered while processing $v_1, \dots, v_{i-1}$. Operation $\mathsf{insert}(i,c)$ inserts interval $(r',i]_c$ into $\mathcal{T}(I_k)$, where $k$ is such that $v_i \in V_k$. Let $x$ be a \emph{relative} pointer to interval $(r',i]_c$ inside $\mathcal{T}(I_k)$. If 
        $\lceil r'/\sigma\rceil \neq \lceil i/\sigma\rceil$ 
        (i.e. $v_{r'}$ and $v_i$ belong to two different blocks), then 
        we doubly-link $P[c]$ and $x$. 
        In any case, we update $P[c]\gets x$ to record that $(r',i]_c$ is the last $c$-labeled interval we have seen.
    \end{definition}

    When processing node $v_i$, we perform the following actions, in order. 
    
    \begin{enumerate}
    \item For every $c \in \mathsf{OUTL}(v_i)$, we call $\mathsf{insert}(i,c)$.
    
    \item If $i\mod \sigma = 0$, i.e. $v_i$ is the last node of its block $V_k$, we call $\mathsf{insert}(i,c)$ for every $c\in\Sigma$.
    \end{enumerate}

    We distinguish two cases in step (2) above: 
    
    (a) If $c \notin \mathsf{OUTL}(v_i)$, then we are inserting in $I_k$ an interval $(r',i]_c$ corresponding to a piece of a $c$-labeled interval of $I$ that continues in block $V_{k+1}$. Since $i$ is the last node in its block, $(r',i]_c$ will be correctly linked to the next piece $(i,j]_c$ of that interval 
    the next time we call $\mathsf{insert}(j,c)$ for some $j>i$.
    
    (b) In the other case ($c \in \mathsf{OUTL}(v_i)$), by step (1) it holds that $P[c]$ is a pointer to an interval of the form $(l,i]_c$, for some $l<i$. Then, step (2) inserts in $I_k$ an \emph{empty} interval $(i,i]_c$. 
    We consider such an interval a dummy interval.
    By Definition \ref{def:insert interval}, note that $(l,i]_c$ and $(i,i]_c$ are not linked together. This is correct, since $(l,i]_c$ is the last piece of a $c$-labeled interval of $I$ (i.e. an interval which does not continue in $V_{k+1}$). On the other hand, the next $c$-labeled interval $(i,r]_c$ we will encounter, for some $r>i$, will be linked with $(i,i]_c$. This is also not an issue: it simply means that some linked lists begin with a (unique) dummy interval, which therefore does not affect the asymptotic cost of interval deletion queries. 
    Note that dummy intervals are empty and thus do not affect stabbing queries as well. 

    After processing the last node $v_n$, for each $c\in \Sigma$ we regard as dummy and delete all intervals contained in the linked list ending in $P[c]$. By construction, this linked list is of the form  $(l,k\sigma] \leftrightarrow (k\sigma,(k+1)\sigma]  \leftrightarrow ((k+1)\sigma,(k+2)\sigma]_c \leftrightarrow  \dots \leftrightarrow ((k+t)\sigma,n]_c$  for some integers $k,t$ (pointer $P[c]$ points to $((k+t)\sigma,n]_c$), where $v_{l}$ is the last node such that $c \in \mathsf{OUTL}(v_l)$. This step is also correct, since those intervals do not correspond to any bridge, precisely because $v_{l}$ is the last node such that $c \in \mathsf{OUTL}(v_l)$.
    Similarly, we will regard as dummy all intervals linked to the initial dummy intervals $(-1,0]_c$, for all $c\in\Sigma$.
    These linked lists are of the form
    $(-1,0] \leftrightarrow (0,k\sigma] \leftrightarrow (k\sigma,(k+1)\sigma]  \leftrightarrow ((k+1)\sigma,(k+2)\sigma]_c \leftrightarrow  \dots \leftrightarrow ((k+t)\sigma,r]_c$ for some integers $k,t$, where $v_{r}$ is the first node such that $c \in \mathsf{OUTL}(v_r)$.
    Also those intervals do not correspond to any bridge and can safely be deleted.

    As noted above,  $\mathcal T(I_k)$ (plus the pointers to adjacent trees' intervals) uses $O(|I_k|\log\sigma)$ bits of space and $|I_k|\in O(\sigma^2)$. 
    Therefore, building each $\mathcal T(I_k)$ takes $O(|I_k|\log|I_k|)\subseteq O(|I_k|\log\sigma)$ time. 
    Let us compute the total number of intervals $|\bigcup_k I_k| = \sum_{k}|I_k|$. Imagine the following process of transforming $I$ into $\bigcup_k I_k$: process the blocks $(k\sigma,(k+1)\sigma]$, for $k=0, 1, \dots$; when processing the $k$-th block, break into two intervals all the intervals of $I$ overlapping both positions $k\sigma$ and $k\sigma+1$ (i.e. intervals overlapping both the $(k-1)$-th and $k$-th blocks). Since no more than $\sigma$ intervals can pairwise intersect, at every processed block the number of intervals in $I$ increases at most by an additive term $\sigma$. Since the number of blocks is $n/\sigma$
    and the original interval set satisfies $|I| \in O(n)$ (this follows from Lemma \ref{lem: bridge-LCP one-to-one}), we obtain that $\sum_{k}|I_k| \leq |I| + \sigma \cdot (n/\sigma) = O(n)$. Although we create some dummy intervals, each $\mathcal T(I_k)$ contains $O(\sigma)$ dummy intervals, which are therefore $O(\sigma)\cdot(n/\sigma) = O(n)$ in total.
    We conclude that, overall, building the $n/\sigma$ interval trees takes $O(n\log\sigma)$ time, and that the interval trees use in total $O(n\log\sigma)$ bits of space.

    It is immediate to see how to perform a query $\mathcal{I}.\mathsf{stab\_and\_remove}(i)$: query the interval tree $\mathcal T(I_k)$ corresponding to block $V_k$ containing node $v_i$, and delete the stabbed intervals. If we stab an interval $(l',r']_c \in I_k$ being a ``piece'' of an interval $(l,r]_c$ in the original set $I$ (i.e. $(l,r]_c$ spans $V_k$ and at least one adjacent block), scan the doubly-linked list associated with $(l',r']_c$ and delete all the intervals (belonging to adjacent blocks $I_{k-1}, I_{k+1}, I_{k-2}, I_{k+2},\dots$)   
    connected by the list (i.e. intervals corresponding to the other ``pieces'' of $(l,r]_c$). 
    
    Since deleting an interval in $\mathcal T(I_k)$ takes constant time given a pointer to it, and an interval of $I$ of length $\ell'$ is broken into $\ell'/\sigma$ pieces (inserted in an interval of sets within $I_0, \dots, I_{n/\sigma-1}$), we obtain our claim: $\mathcal{I}.\mathsf{stab\_and\_remove}(i)$ is solved in $O(\log \sigma + |R| + \ell/\sigma)$ time, where $|R|$ is the number of stabbed intervals and $\ell$ is the cumulative length of the stabbed intervals. 

\end{proof}

\end{document}